\documentclass[12 pt]{article}
\usepackage{amssymb}
\usepackage{amsmath}
\usepackage{hyperref}
\usepackage{graphicx}
\usepackage{placeins}
\usepackage[font=small,labelfont=bf]{caption}
\usepackage{bm}
\usepackage{latexsym}
\begin{document}
\title{\bf Exponential Modification of AdS Black Hole and Thermodynamic Behavior}
\author{  Mehdi Sadeghi\thanks{Corresponding author: Email:  mehdi.sadeghi@abru.ac.ir}\,\,\,  and  \,\,Faramaz Rahmani\thanks{Email:  faramarz.rahmani@abru.ac.ir}\hspace{2mm}\\
	{\small {\em Department of Physics, Faculty of Basic Sciences,}}\\
	{\small {\em Ayatollah Boroujerdi University, Boroujerd, Iran}}
}
\date{\today}
\maketitle

\abstract{ In this paper, we present an exponential modification for the action of an AdS black hole in the absence of a matter field. An approximated black hole solution is obtained up to the third order of perturbation coefficient. A thermodynamic investigation in canonical ensemble shows that the behavior of a Van der Waals fluid is not seen in this model. Nevertheless, the study of thermodynamic potentials and other related quantities suggests that the thermodynamic phase transitions of the first and second types can occur in this model. The forms of the phase transitions are more similar to the Hawking-Page phase transitions.}\\

%\pacs{11.10.Jj, 11.10.Wx, 11.15.Pg, 11.25.Tq}


%--------------------------------------------------------------------------
\section{Introduction} \label{intro}
Einstein's theory of general relativity has been one of the most successful theories of physics so far. Before this theory, the nature of gravity was not obvious to anyone. But with Einstein's clever idea, which originated from the principle of equivalence, gravity was introduced as the curvature of spacetime, and experimental observations after that confirmed the correctness of this theory. One of the challenging issues has been the issue of unifying gravity and quantum mechanics. So far, such an approach suffers from several problems such as renormalization of the theory or the presence of ghosts in the solutions of the theory. It is expected that the modification of gravitational actions with higher order terms helps us to create a quantum theory of gravity. Therefore, during the last few decades, many efforts have been made in this field. Nevertheless, general relativity (GR), which was formulated by Albert Einstein in 1915, has achieved significant milestones such as gravitational waves, the Mercury anomaly, gravitational lensing, and black holes. Observational data confirm that 95$\%$ of the Universe consists of dark energy (about 69$\%$) and dark matter (about 26$\%$), which are unknown. GR has failed to account for this unidentified portion of the Universe. Additionally, General Relativity's non-renormalizability has impeded its quantization using conventional methods in quantum field theory. Therefore, modifications were made to GR to address these limitations, including the incorporation of Dark Energy and Dark Matter, and the unification of Gravity with quantum mechanics\cite{Stelle:1976gc}. $F(R)$ gravity\cite{Brans:1961sx},\cite{Hu:2007nk}\cite{Nojiri:2003ft}, Lovelock gravity\cite{Lovelock:1971yv}, Gauss-Bonnet gravity\cite{Nojiri:2005jg}\cite{Sadeghi:2015vaa}, Quasi-topological gravity\cite{Parvizi:2017boc}, massive gravity\cite{Fierz:1939ix},\cite{deRham:2014zqa},\cite{Sadeghi:2018ylh} the Rastall theory of gravity\cite{Rastall:1972swe},\cite{Sadeghi:2023tzf} and exponential modified gravity\cite{Kruglov:2012cq}\cite{Cognola:2007zu}\cite{Nojiri:2010wj}\cite{Nojiri:2017ncd} are some of the modifications of general relativity found in the literature.\par
One of the places where quantum and gravitational concepts can intertwine and make us understand more relevant concepts is where the thermodynamics of black holes are investigated. As we know, black holes have thermodynamic properties as solutions to the theory of general relativity. Black hole radiation (Hawking radiation) is a quantum phenomenon, and where gravity is very strong, a quantum gravity theory is required to properly understand the physics of this phenomenon. Therefore, the thermodynamic study of the black hole can be of great importance. Specifically, the study of phase transition in black holes provides insights into the fundamental properties of black holes and enables us to explore their thermodynamic stability. The thermodynamic behavior of black holes can be translated by using the AdS/CFT dictionary to study some fundamental issues like confinement-deconfinement phase transition or the problem of superconductivity which are investigated in the context of a conformal quantum field theory\cite{Maldacena:1997re}\cite{Aharony:1999ti}\cite{Dey:2015poa}. Phase transitions in black holes may open a way toward a quantum theory of gravity. They offer a playground for testing theories that attempt to unify general relativity and quantum mechanics, such as string theory and loop quantum gravity. The investigation of black hole thermodynamics and their phase transitions helps us to realize the quantum nature and fundamental laws governing black holes. In summary, phase transitions in black hole thermodynamics can provide profound insights into the nature of black holes, the structure of spacetime and the fundamental laws of physics.\par  
In section (\ref{sec2}), we introduce the exponential modification of the AdS black hole. Then, we obtain an approximated black hole solution up to the third order of perturbation coefficient. Finally, in section (\ref{sec3}), the thermodynamic behavior of the model shall be studied in canonical ensemble to determine the types of phase transitions and stability of the system. 
%--------------------------------------------------------------------------
\section{ Exponential Modification of the AdS Black Hole}
\label{sec2}
The 4-dimensional action of exponentially modified gravity with negative cosmological constant is \cite{Kruglov:2012cq},
\begin{eqnarray}\label{action}
S=\frac{1}{16\pi G}\int d^{4}  x\sqrt{-g} \bigg[R e^{q R}-2\Lambda\bigg],
\end{eqnarray}
where $R$ is the Ricci scalar, $\Lambda=-\frac{3}{l^2}$ is the cosmological constant and $l$ is the AdS radius. When $q \to 0 $, the exponential term gets transformed into standard Einstein-Hilbert gravity.
The equations of motion are obtained by varying action (\ref{action}) with respect to  $g_{\alpha \beta } $. This leads to equation
\begin{align}\label{EH}
	&  e^{q R} R_{\alpha  
		\beta  }+\Lambda  g_{\alpha  \beta  }  - \tfrac{1}{2} e^{q R} g_{\alpha  \beta  } R + e^{q R} q R_{\alpha  \beta  } R
	-3 q^2 e^{q R}  \nabla _{\alpha  }R \nabla _{\beta  \
	}R - e^{q R} q^3 R \nabla _{\alpha  }R \nabla_{\beta  }R \nonumber\\& -2 q e^{q R}  \nabla _{\beta  }\nabla _{\alpha  }R - q^2 e^{q R}  R \nabla _{\beta  }\nabla _{\alpha  }R + 2 q e^{q R}  g_{\alpha  \beta  } \nabla _{\gamma  }\nabla ^{\gamma  }R +q^2 e^{q R} g_{\alpha  \beta  } R \nabla \
	_{\gamma  }\nabla ^{\gamma  }R \nonumber\\&+ 3 q^3 e^{q R} q^2 g_{\alpha  \beta  } \nabla _{\gamma  }R \nabla ^{\gamma  }R + e^{q R}  g_{\alpha  \beta  } R \nabla
	_{\gamma  }R \nabla ^{\gamma  }R=0.
\end{align}
Our goal is to find an asymptotically AdS black hole solution of the above equation in four-dimensional maximally symmetric spacetime. In this regard, we consider the following ansatz,
\begin{equation}\label{metric1}
ds^{2} =-f(r)dt^{2} +\frac{dr^{2} }{f(r)} +r^2d\theta^2+r^2 \sin^2 \theta d\phi^2.
\end{equation}\\
Where $f(r)$ is the metric function that shall be determined.\\
By using ansatz (\ref{metric1}) and relation (\ref{EH}), the $tt$ component of Eq. (\ref{EH}) gives 
\begin{equation}\label{eom}
\begin{split}
&64 q^3 f(r)^4-16 q^2 f(r)^3 \left(-12 q r f'(r)+6 q r^2 f''(r)+2 q r^3 f^{(3)}(r)+12 q+7 r^2\right)\\&-r^6 \left(2 \Lambda  r^2 \exp \left(\frac{q \left(4 r f'(r)+r^2 f''(r)+2 f(r)-2\right)}{r^2}\right)+q r^2
f''(r)^2-2 q f''(r)-2\right)\\&-r^3 f'(r) \left(4 q^2 r^4 f''(r)^2-2 \left(q r^3 f^{(3)}(r) \left(q+r^2\right)+4 q^2+6 q
r^2-r^4\right)\right)+8 q^2 r^5 f'(r)^3 \\&-q r^5 f'(r) f''(r)\left(q r^3 f^{(3)}(r)-2 \left(2 q+r^2\right)\right)\\&-2 q r^4 f'(r)^2
\left(7 q r^2 f''(r)+2 q r^3 f^{(3)}(r)+10 q+6 r^2\right)\\&
+4 q f(r)^2 \left(48 q^2+56 q r^2+4 r^4+36 q^2 r^2 f'(r)^2+16 q^2 r^3 f^{(3)}(r)+6q r^5 f^{(3)}(r)+q^2 r^6 f^{(3)}(r)^2\right)\\&+16 r^2 q^2 f(r)^2 f''(r) \left(q r^3 f^{(3)}(r)+12 q+11 r^2\right)\\&-2 q r f'(r) \left(36 q r^2 f''(r)+10 q r^3
	f^{(3)}(r)+48 q+13 r^2\right)\\&+2 f(r) \left(16 q^3 r^3 f'(r)^3+16 q^3 r^6 f''(r)^3-12 q^2 r^5 f^{(3)}(r)-16 q^3 r^3 f^{(3)}(r)-56 q^2 r^2-32 q^3-8 q
	r^4\right)\\&+2 f(r) \left(2 q^2 r^6 \left(4 q r f^{(3)}(r)-25\right) f''(r)^2-3 q^2 r^8 f^{(3)}(r)^2-2 q^3 r^6 f^{(3)}(r)^2+12 q r^7
	f^{(3)}(r)-r^6\right)\\&+2 f(r) \left(-2 q^2 r^2 f'(r)^2 \left(30 q r^2 f''(r)+8 q r^3 f^{(3)}(r)+36 q-7 r^2\right)+2 q^2 r^6 f^{(4)}(r)+2 q r^8
	f^{(4)}(r)\right)\\&
	-2 q r f(r) f'(r) \left(48 q^2 r^4 f''(r)^2+q r^3 f^{(3)}(r) \left(40 q-13 r^2\right)+96 q^2+48 q r^2-14 r^4\right)\\&-2 q r f(r) f'(r) \left(4 q^2 r^6 f^{(3)}(r)^2+2 q r^2 \left(14 q r^3 f^{(3)}(r)+72 q+21 r^2\right) f''(r)-4 q r^6
	f^{(4)}(r)\right)\\&
	+2 q r^2 f(r) \left(q^2 r^6 f^{(3)}(r)^2-q r^6 f^{(4)}(r)-2 q r^3 f^{(3)}(r) \left(4 q+15 r^2\right)\right) f''(r) \\&
	-2 q r^2 f(r) \left(48 q^2+88 q r^2-3
	r^4\right) f''(r)=0.
\end{split}
\end{equation}
An exact solution for $f(r)$ cannot be obtained. So, we consider a perturbed form for $f(r)$ up to the third order of $q$ as bellow,
\begin{equation}\label{fapp}
	f(r)=f_0(r)+q f_1(r)+q^2 f_2(r)+q^3 f_3(r),
\end{equation}
By substituting the relation (\ref{fapp}) into equation (\ref{eom}) and considering  the coefficient of the zeroth order of $q$, the following equation is obtained as bellow,
\begin{equation}
r f_0'(r)+f_0(r)+\Lambda  r^2-1=0.
\end{equation}
The function $f_0(r)$ is easily obtained through the solving of the above equation. The result is
\begin{equation}
	f_0(r)= 1+\frac{C_1}{r}-\frac{\Lambda r^2}{3}.
\end{equation}
Where, $C_1$ is an integration constant related to the mass of the black hole.\par 
%%%
Up to the first order of $q$ is, Eq. (\ref{eom}) gives
\begin{equation}
\begin{split}
&	2 r^7 f_0'(r f_0''(r)+2 r^8 f_0^{(3)}(r) f_0'(r)-12 r^6 f_0'(r)^2-28 r^5 f_0(r)
	f_0'(r)\\&+12 r^5 f_0'(r)+r^8 \left(-f_0''(r)^2\right)+6 r^6 f_0(r) f_0''(r)+2 r^6
	f_0''(r)+4 r^8 f_0(r) f_0^{(4)}(r)\\&+24 r^7 f_0(r) f_0^{(3)}(r)+16 r^4 f_0(r)^2-16 r^4
	f_0(r)-2 r^7 f_1'(r)-2 r^6 f_1(r)=0.
\end{split}
\end{equation}
This leads to $f_1(r)$ as follows,
\begin{equation}
f_1(r)=\frac{C_2}{r}+\frac{1}{r}\int^r {\frac{A(u)}{2 u^2}},
\end{equation}
where $C_2$ is a constant and $A(u)$ has the form
\begin{align}
	A(u)&=-f_0''(u)^2 u^4+2 f_0'(u) f_0^{(3)}(u) u^4+4 f_0(u) f_0^{(4)}(u)
	r^4+2 f_0'(u) f_0''(u) u^3\nonumber\\&+24 f_0(u) f_0^{(3)}(u) u^3-12 f_0'(u)^2 u^2+6
	f_0(u) f_0''(u) u^2+2 f_0''(u) u^2\nonumber\\&-28 f_0(u) f_0'(u) u+12 f_0'(u) u+16
	f_0(u)^2-16 f_0(u).
	\end{align}
Finally, $f_1(r)$ takes the form
\begin{equation}
	f_1(r)=\frac{C_2}{r}-\frac{4 \Lambda^2}{3}r^2.
\end{equation}
Up to the second order of $q$ in Eq.(\ref{eom}), the related coefficient gives an equation as follows,
\begin{equation}
\begin{split}
&r^3 \left(-24 r^3 f_0'(r) f_1'(r)-2 r^5 f_0''(r) f_1''(r)+8 r^2 f_0'(r)^3+2 r^3
	f_1''(r)-2 r^4 f_2'(r)\right)-112 r^2 f_0(r)^3 \\&-\Lambda  r^4 \left(\left(4 r f_0'(r)+r^2 f_0''(r)+2 f_0(r)-2\right)^2+2 r^2 \left(r \left(4
	f_1'(r)+r f_1''(r)\right)+2 f_1(r)\right)\right)\\&-2 r^4 f_0'(r)^2 \left(7 r^2 f_0''(r)+2 r^3 f_0^{(3)}(r)+10\right)+2 r^5 \left(r^2 \left(f_0''(r)+r f_0^{(3)}(r)\right)+6\right) f_1'(r)\\&-r^3 f_0'(r) \left(-2 r^3 \left(f_0^{(3)}(r)+r f_1''(r)\right)+4 r^4 f_0''(r)^2+r^2 \left(r^3
	f_0^{(3)}(r)-4\right) f_0''(r)-8\right)\\&+4 f_0(r) \left(r^2 f_0(r) \left(-26 r f_0'(r)+44 r^2 f_0''(r)+r^4 \left(-f_0^{(4)}(r)\right)+6
	r^3 f_0^{(3)}(r)+56\right)+8 r^4 f_1(r)\right)\\&+2 r^4 f_1(r) \left(r \left(-14 f_0'(r)+3 r \left(f_0''(r)+4 r f_0^{(3)}(r)\right)+2 r^3
	f_0^{(4)}(r)\right)-8\right)-2 r^6 f_2(r)\\&+2 r^3 f_0(r) \left(f_0'(r) \left(42 r^2 f_0''(r)-13 r^3 f_0^{(3)}(r)+48\right)+14 r
	f_0'(r)^2\right)-112 r^2 f_0(r)\\&-2 r^3 f_0(r) \left(r \left(50 r f_0''(r)^2+6 f_0^{(3)}(r) \left(5 r^2 f_0''(r)+2\right)+3 r^3
	f_0^{(3)}(r)^2+14 f_1'(r)-3 r f_1''(r)\right)+88 f_0''(r)\right)\\&-2 r^6 f_0(r) f_0^{(4)}(r) \left(4 r f_0'(r)+r^2 f_0''(r)-2\right)=0,
\end{split}
\end{equation}
By solving this equation, $f_2(r)$ is given by
\begin{equation}
\begin{split}
f_2(r)=&\frac{C_3}{r}-\frac{8 \Lambda ^3 r^3}{r}-\frac{6 C_1 C_2}{r^4}-\frac{12 C_2}{r^3}\\
&+\frac{32 \Lambda ^2 r}{3 r}+\frac{16 C_1 \Lambda ^2 \ln
		(r)}{r}-\frac{32 \Lambda }{3}\&-\frac{16 C_1 \Lambda  \ln (r)}{r}-\frac{12 C_2 \Lambda  \ln (r)}{r}.
\end{split}
\end{equation}
As the same way, $f_3(r)$ is obtained as follows,
\begin{equation}
\begin{split}
f_3(r)=&\frac{C_4}{r}+\frac{128 \Lambda ^3}{3}-\frac{128 \Lambda ^2}{3}-\frac{2080}{27} \Lambda ^4 r^2-\frac{352 \Lambda ^3 r^2}{27}+\frac{128 \Lambda ^2
		r^2}{27}-\frac{256 \Lambda ^2}{3 r^2}\\&-\frac{80 C_1^2 \Lambda ^2}{r^4}-\frac{160 C_1 \Lambda ^2}{r^3}+\frac{256 \Lambda }{3 r^2}+\frac{80
		C_1^2 \Lambda }{r^4}+\frac{160 C_1 \Lambda }{r^3}-\frac{84 C_1 C_2 \Lambda }{r^4}+\frac{48 C_2 \Lambda }{r^3}\\&-\frac{264 C_2 \Lambda
	}{r^4}-\frac{6 C_2{}^2}{r^4}+\frac{396 C_1^2 C_2}{r^7}+\frac{432 C_1 C_2}{r^6}+\frac{792 C_1 C_2}{r^7}+\frac{864 C_2}{r^6}\\&-\frac{16 C_1
	\Lambda ^3 \ln (r)}{r}+\frac{16 C_1 \Lambda ^2 \ln (r)}{r}+\frac{12 C_2 \Lambda ^2 \ln (r)}{r}-\frac{48 C_2 \Lambda  \ln
		(r)}{r}.
\end{split}
\end{equation}
 \begin{figure}[h!]
\centering
[a]{\includegraphics[width=7cm]{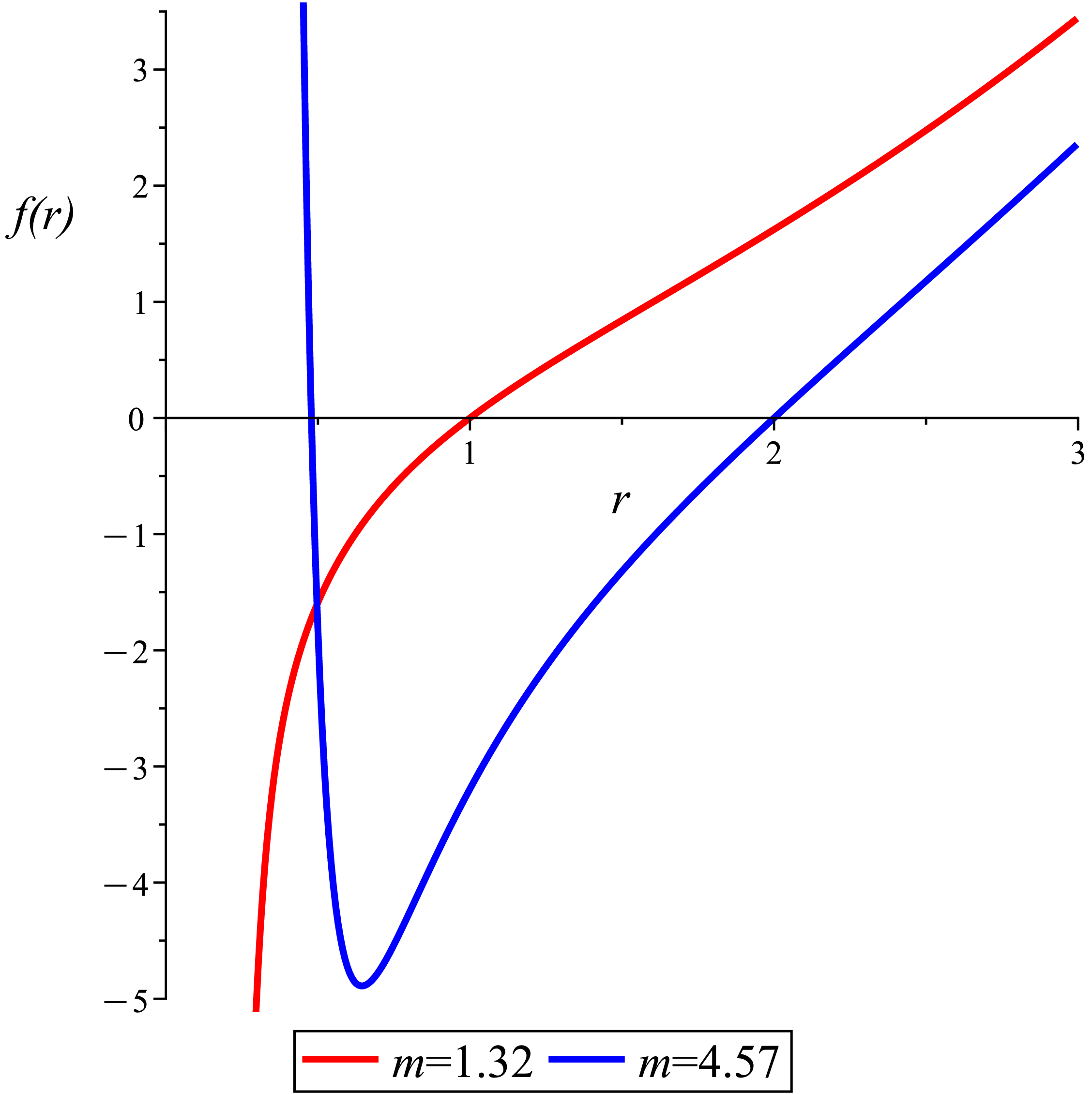}}\quad
[b]{\includegraphics[width=7cm]{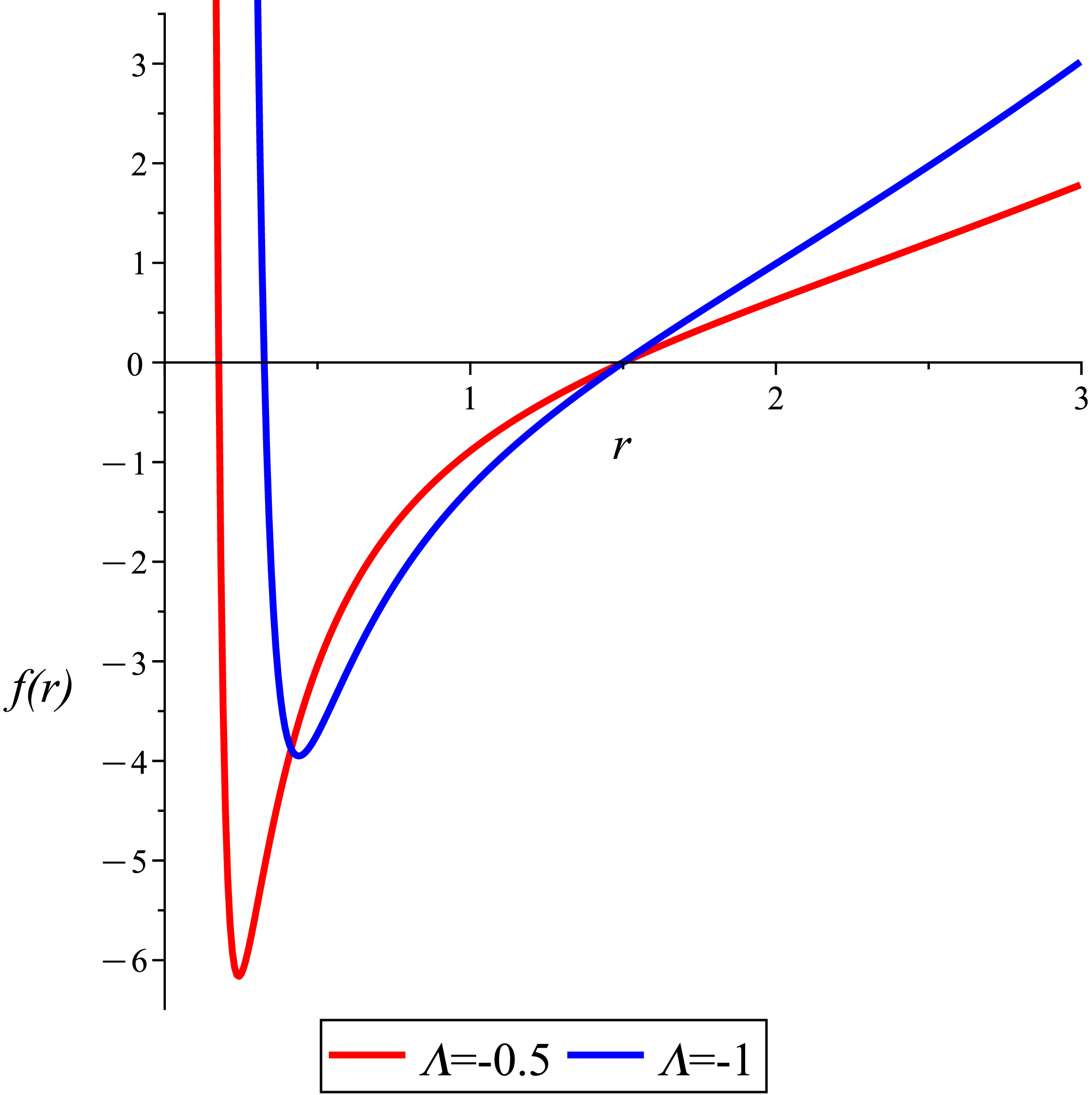}}
\caption{$f(r)-r_h$ diagrams when \textbf{(a)}: $q=0.01$, $\Lambda=-1$; \textbf{(b)}: $q=0.01$ and  outer horizon is at $r_h=1.5$. In the left panel the value of the $\Lambda$ is fixed and by increasing the value of $m$, the number of roots increases. In the right panel the value of  $\Lambda$ changes. \label{fig:f1}}
\end{figure}	
Since, each of the functions $f_0, f_1, f_2$ and $f_3$ vanish on the horizon of the black hole, the unknown coefficients $C_1, C_2, C_3$ and $C_4$ can be determined easily. The numerator of the term $\frac{2m}{r^{d-3}}$, in metric functions is proportional to the integration constant $m$. With this consideration, the ADM mass is obtained in terms of powers of $q$ which up to the second order of $q$ is expressed as bellow,
\begin{equation}\label{ma}
m=m^{(0)}+m^{(1)}+m^{(2)}+\cdots,
\end{equation}
where,
\begin{equation}\label{m0}
m^{(0)}=\frac{r_{h}}{2}-\frac{\Lambda  r_{h}^{3}}{6},
\end{equation}
\begin{equation}\label{m1}
m^{(1)}=-\frac{2 \Lambda^{2} r_{h}^{3} q}{3},
\end{equation}
and
\begin{equation}\label{m2}
\begin{split}
m^{(2)}&=\Big( 4 r_{h} \Lambda^{2} -8 r_{h} \Lambda^{2} \ln(r_{h}) +8 r_{h} \Lambda  \ln (r_{h}) 	-\frac{8 r_{h}^{3} \Lambda^{2} \ln  (r_{h})}{3}-\frac{16 r_{h}^{3} \Lambda^{3} \ln(r_{h})}{3}\\
-&\frac{16 r_{h}^{3} \Lambda^{3}}{3}-\frac{16 r_{h} \Lambda}{3}-\frac{8 r_{h} \Lambda^{2}}{3}\Big) q^{2}.
\end{split}
\end{equation}
The behavior of $f(r)$ for various values of parameters is seen in Fig. (\ref{fig:f1}). Investigations show that for different values of parameters, the function $f(r)$ has at most two roots in this model.\par 
Now, the Hawking temperature\cite{Hawking:1974rv} for this black hole solution up to the second order of $q$ is
  \begin{equation}\label{Temp}
  \begin{split}
  T =&\frac{f'(r_h)}{4 \pi}=\frac{1}{4 r_{h} \pi}-\frac{r_{h} \Lambda}{4 \pi}-\frac{r_{h} \Lambda^{2}}{\pi}q+ \left(-\frac{20 r_{h} \Lambda^{3}}{3 \pi}-\frac{4 r_{h} \Lambda^{2}}{3 \pi}+\frac{2 \Lambda^{2}}{3 r_{h} \pi}+\frac{4 \Lambda}{3 r_{h} \pi}\right) q^{2} +O(q^3).
  \end{split}
  \end{equation}
One can see that when $q=0$, the mass and temperature of this model are the same as the mass and temperature of the Schwarzchild-AdS black hole.\par 
The entropy of our model is calculated using the Wald formula \cite{Wald:1993nt} as follows,
\begin{equation}\label{Wald}
	S= 2 \pi \int_{\text{Horizon}} d^2x \sqrt{-g} \, \, \, \epsilon_{\alpha \beta } \epsilon_{\rho \gamma }\,\,\frac{\partial \mathcal{L}}{R_{\alpha \beta \rho \gamma}},
\end{equation}
where,
\begin{eqnarray}
	\epsilon_{\alpha \beta } \epsilon_{\rho \gamma }\frac{\partial \mathcal{L}}{R_{\alpha \beta \rho \gamma}}= \frac{1}{16 \pi G}\epsilon_{\alpha \beta } \epsilon^{\alpha \beta } (1+q R) e^{q R}.
\end{eqnarray}
Furthermore,
\begin{equation}
	\epsilon_{\alpha \beta}= \delta^{t}_{\alpha}\delta^{r}_{\beta}-\delta^{t}_{\beta}\delta^{r}_{\alpha},
\end{equation}
\begin{equation}
	\epsilon_{01}=-\epsilon_{10}=1 ,
\end{equation}
and
\begin{equation}
	\epsilon_{\alpha \beta } \epsilon^{\alpha \beta }=2.
\end{equation}
Calculations give the following relation for the Ricci scalar, 
\begin{equation}\label{rr}
	R(r)=-\frac{4 r f'(r)+r^2 f''(r)+2 f(r)-2}{r^2},
\end{equation}
where, at the black hole horizon takes the form
\begin{equation}\label{r3}
	R(r_h)=-\frac{4 r_h f'(r_h)+r_h^2 f''(r_h)-2}{r_h^2},
\end{equation}
because, $f(r_h)=0$. 
Hence, Eq. (\ref{Wald}) leads to
\begin{equation}\label{ss}
	S=\pi r_h^2\left(1+q R(r_h)\right) e^{q R(r_h)}.
\end{equation}
Where after substituting relation (\ref{r3}) into relation (\ref{ss}), we get
\begin{equation}\label{ss2}
S = S^{(0)}+S^{(1)}+S^{(2)}+S^{(3)}+S^{(4)}+\cdots,
\end{equation}
where,
\begin{equation}
S^{(0)}=\pi r_h^2,
\end{equation}
\begin{equation}
S^{(1)}=8 \pi \Lambda r_h^2 q,
\end{equation}
\begin{equation}
S^{(2)}=56 \pi r_h^2 \Lambda^2 q^2,
\end{equation}
\begin{equation}
S^{(3)}=\left(480 \pi  r_h^{2} \Lambda^{3}+\frac{32}{3} \pi  r_h^{2} \Lambda^{2}-\frac{128}{3} \Lambda^{2} \pi +\frac{32}{3} \pi  \Lambda \right) q^{3},
\end{equation}
and
\begin{equation}
\begin{split}
S^{(4)}=&\Big(\frac{12704 \pi  r_{h}^{2} \Lambda^{4}}{9}+\frac{3488 \pi  r_{h}^{2} \Lambda^{3}}{9}-\frac{1024 \pi  r_{h}^{2} \Lambda^{2}}{9}+\frac{32864 \pi  \,\Lambda^{3}}{3}\\
-&\frac{16896 \pi  \,\Lambda^{3}}{r_{h}}+\frac{2080 \Lambda^{2} \pi}{3}-\frac{8320 \Lambda^{2} \pi}{r_{h}^{2}}+\frac{17280 \Lambda^{2} \pi}{r_{h}^{3}}-\frac{320 \pi  \Lambda}{r_{h}^{2}}\Big) q^{4}.
\end{split}
\end{equation}
\begin{figure}[h!]
\centering
[a]{\includegraphics[width=7cm]{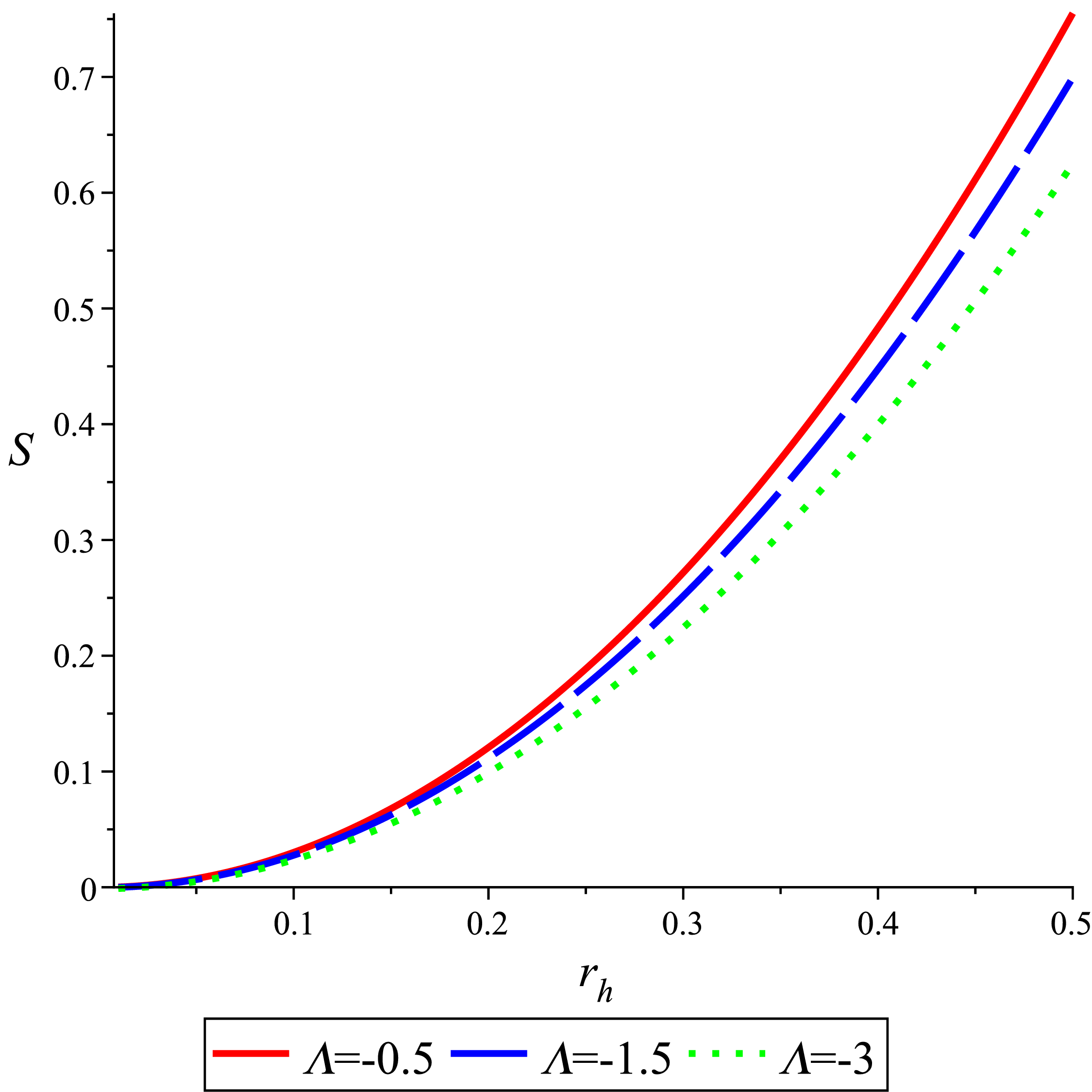}}\quad
[b]{\includegraphics[width=7cm]{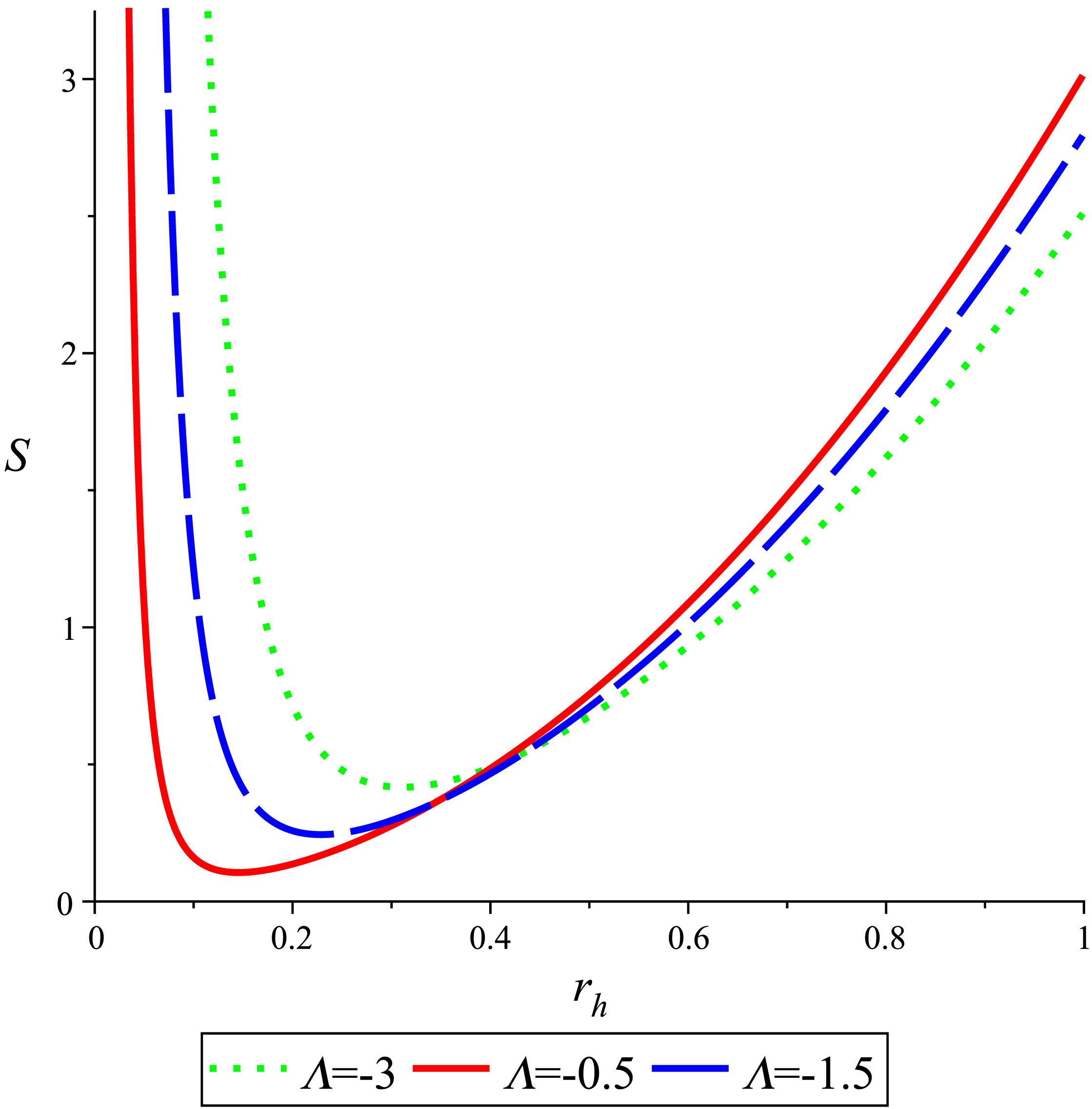}}
\caption{The diagrams of entropy versus the black hole horizon for several values of $\Lambda$ and $q=0.01$; \textbf{(a)}: up to the third order of $q$ and  \textbf{(b)}: up to the fourth order of $q$. In the right panel, for small values of the black hole horizon, the entropy of the system increases. \label{fig:entropy}}
\end{figure}	
As we expect, by increasing the horizon of the black hole, the entropy of the system  increases and vice versa. This is obvious from the left panel of Fig.(\ref{fig:entropy}) which shows the entropy diagram up to the third order of $q$. But, in the right panel which shows the entropy up to the fourth order of $q$, a minimum for each diagram is seen and for small horizons, the entropy of the system increases. This is not a classical phenomenon and may be a sign of quantum effects.
%%%%%%%%%%%%%%%%%%%%%%%%%%%%%%%%%
\section{Thermodynamic Stability and Phase Transitions}
\label{sec3}
Usually, in the study of thermodynamic behavior of black hole solutions in extended phase space, we need to obtain the mass of the black hole which plays the role of enthalpy of the system. For a four-dimensional AdS black hole, the enthalpy $M$ is equal to the integration constant $m$ which appears in the numerator of the term $\frac{2m}{r^{(d-3)}}$  in the metric function. Here, $d$ stands for the dimension of spacetime. The general relation between the enthalphy (mass) of the black hole and integration constant $m$ is as follows\cite{Dolan:2010ha},
\begin{equation}\label{ent}
M=\frac{(d-2)\Omega_{(d-2)}}{8 \pi }m,
\end{equation}
where, $\Omega_{(d-2)}$ is the volume of the $d-2$-dimensional unit sphere and is determined through the relation
\begin{equation}\label{surface}
\Omega_{(d-2)}=\frac{2 \pi^{\frac{d-2}{2}}}{\Gamma(\frac{d-2}{2})}.
\end{equation}
In $4$-dimensional models, the enthalpy of the system is equal to the integration constant $m$. Hence, by using the relations (\ref{m0}), (\ref{m1}) and (\ref{m2}) and $\Lambda=-\frac{P}{8 \pi}$\cite{Dolan:2010ha}, the enthalpy of the system is given by
 \begin{equation}\label{ent}
  \begin{split}
  M(P,r_h,q)=&\frac{ r_{h}}{2}+\frac{4 \pi  P r_{h}^{3}}{3}-\Big(\frac{128 \pi^{2} P^{2} r_{h}^{3}}{3} \Big)q\\
  +&\Big( r_{h} \pi^{2} P^{2} -512 r_{h} \pi^{2} P^{2} \ln \! \left(r_{h}\right) -64 r_{h} \pi  P \ln \! \left(r_{h}\right)\\
      -&\frac{512 r_{h}^{3} \pi^{2} P^{2} \ln \! \left(r_{h}\right)}{3}+\frac{8192 r_{h}^{3} \pi^{3} P^{3} \ln \! \left(r_{h}\right)}{3}+ \frac{8192 r_{h}^{3} \pi^{3} P^{3}}{3}\\
  +& \frac{128 r_{h} \pi  P}{3}-\frac{512 r_{h} \pi^{2} P^{2}}{3}\Big) q^{2}+O(q^3).
  \end{split}
  \end{equation}
 The conjugate volume for the system pressure is given by
\begin{equation}\label{vol}
  \begin{split}
 V&=\left(\frac{\partial M}{\partial P}\right)_{r_h,q}=\frac{4 \pi  r_{h}^{3}}{3}-\Big(\frac{256 \pi^{2} P r_{h}^{3} }{3}\Big)q \\
 &+\Big(512 r_{h} \pi^{2} P  -1024 r_{h} \pi^{2} P \ln \! \left(r_{h}\right) -64 r_{h} \pi  \ln \! \left(r_{h}\right)  -\frac{1024 r_{h}^{3} \pi^{2} P \ln \! \left(r_{h}\right)}{3}\\
 &+ 8192 r_{h}^{3} \pi^{3} P^{2} \ln \! \left(r_{h}\right)+
  8192 r_{h}^{3} \pi^{3} P^{2}+\frac{128 r_{h} \pi}{3}-\frac{1024 r_{h} \pi^{2} P}{3}\Big) q^{2}+O(q^3).
  \end{split}
  \end{equation}
When $q=0$, the conjugate volume is equal to the usual volume of a simple  Schwarzschild-AdS black hole.\par  The pressure of this model up to the second order of $q$ is given by,
\begin{equation}
\begin{split}
P &= \frac{T}{2 r_{h}}-\frac{1}{8 \pi  r_{h}^{2}}+\left(\frac{8 \pi  T^{2}}{r_{h}^{2}}-\frac{4 T}{r_{h}^{3}}+\frac{1}{2 \pi  r_{h}^{4}}\right) q \\&+\left(\frac{256 \pi^{2} T^{3}}{r_{h}^{3}}-\frac{192 \pi  T^{2}}{r_{h}^{4}}+\frac{48 T}{r_{h}^{5}}-\frac{4}{\pi  r_{h}^{6}}\right) q^{2}+O(q^3).
\end{split}
\end{equation}
 To have a correct Smarr formula, the coupling constant $q$ should be considered as a charge with a potential$\mathcal{Q}$ as follows,
   \begin{equation}\label{Q}
  \begin{split}
\mathcal{Q}&=\left(\frac{\partial M}{\partial q}\right)_{r_h,P}=-\frac{128 r_{h}^{3} \pi^{2} P^{2}}{3}+ \Big(256 r_{h} \pi^{2} P^{2} -512 r_{h} \pi^{2} P^{2} \ln \! \left(r_{h}\right) -64 r_{h} \pi  P \ln \! \left(r_{h}\right) \\ 
& -\frac{512 r_{h}^{3} \pi^{2} P^{2} \ln \! \left(r_{h}\right)}{3}+
\frac{8192 r_{h}^{3} \pi^{3} P^{3} \ln \! \left(r_{h}\right)}{3}+\frac{8192 r_{h}^{3} \pi^{3} P^{3}}{3}+\frac{128 r_{h} \pi  P}{3}-\frac{512 r_{h} \pi^{2} P^{2}}{3}\Big) q\\&+O(q^2).
  \end{split}
  \end{equation}
Now, the first law of thermodynamics takes the form
  \begin{equation}
  \delta M=T \delta S + V \delta P +\mathcal{Q} \delta q.
  \end{equation}
The Smarr formula reads as\footnote{First, a general $4D$ Smarr formula was considered in the form $M=2 TS-2PV+ x \mathcal{Q}q$. Then, through an examination we realized that the value $x=1$ ensures a correct Smarr formula up to the first order of $q$.}
  \begin{equation}
  M=2 TS-2PV+ \mathcal{Q}q,
  \end{equation}
up to the linear order in $q$.\par
\begin{figure}[h!]
\centering
[a]{\includegraphics[width=7cm]{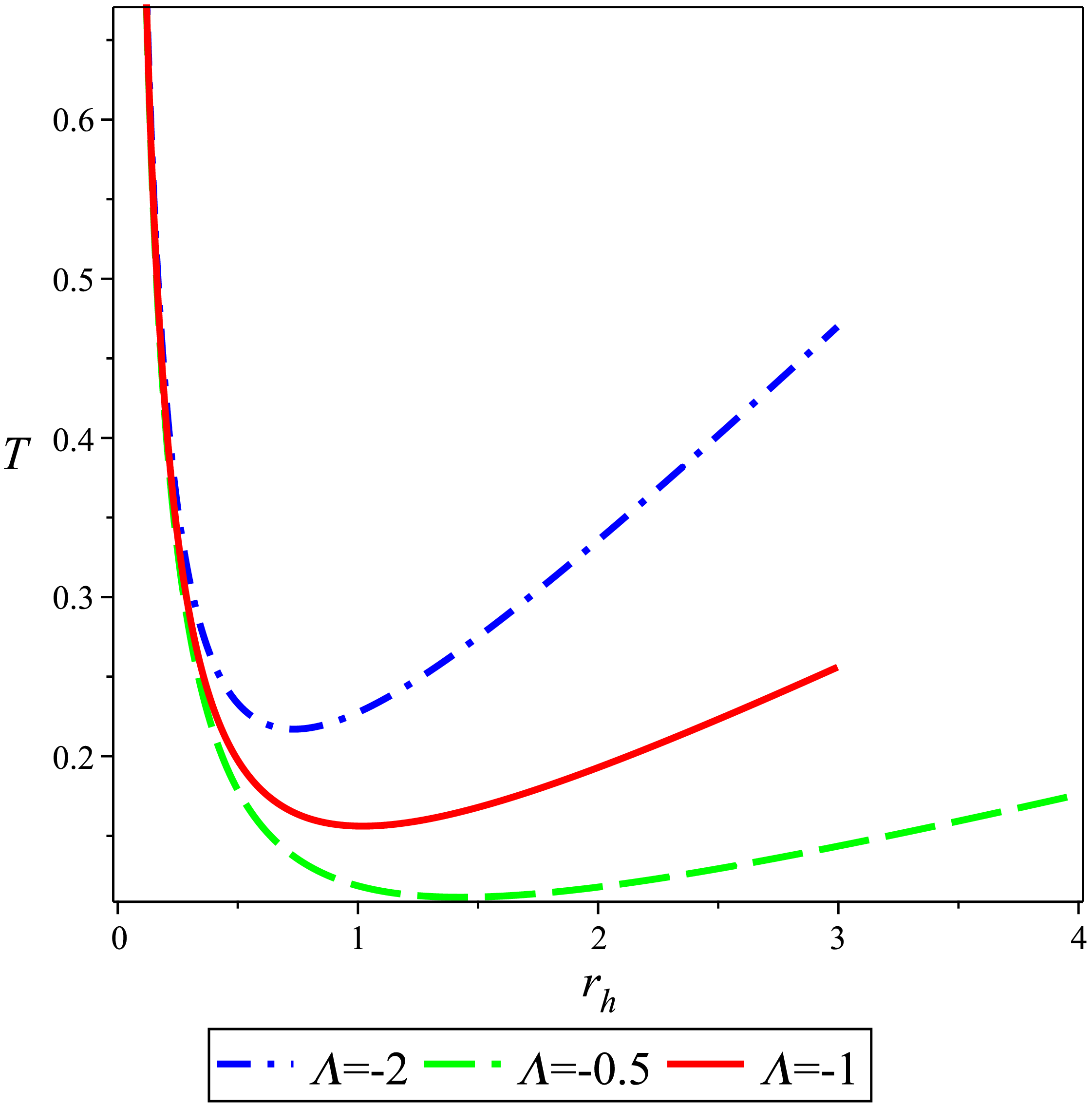}}\quad
[b]{\includegraphics[width=7cm]{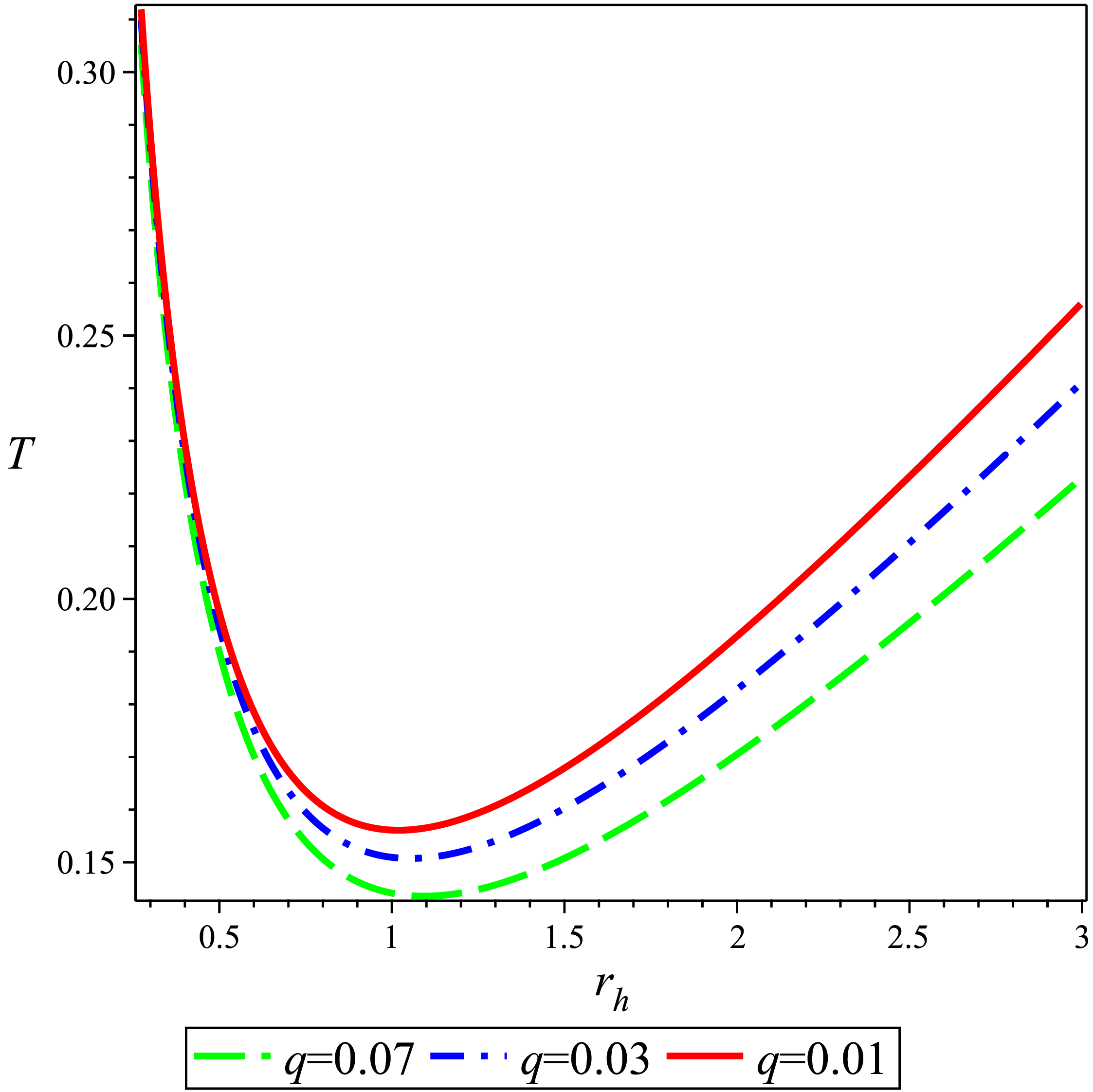}}
\caption{$T-r_h$ diagrams up to the second order of $q$ when \textbf{(a)}: the parameter $\Lambda$ changes and  \textbf{(b)}: the parameter $q$ changes. In both panels,  minimum values are seen for the system temperature. Bellow the minimum values no black hole can exist. Above the minimum value two unstable and stable phases are seen. \label{fig:temp}}
\end{figure}

 
Investigations show that the pressure and temperature diagrams of this model do not correspond to the pressure and temperature diagrams of a Van der Waals fluid. Fig. (\ref{fig:temp}) shows that for different values of parameters, the temperature of the system has a minimum value. The minimum of the temperature when has expressed up to the first and second order of $q$ occurs at horizons
\begin{equation}\label{r01}
r_{\textit{min}}^{(1)}=\frac{1}{\sqrt{-4 \Lambda^{2} q -\Lambda}},
\end{equation}
and
\begin{equation}\label{r02}
r_{\textit{min}}^{(2)}=\frac{\sqrt{-\Lambda  \left(80 \Lambda^{2} q^{2}+16 \Lambda  q^{2}+12 \Lambda  q +3\right) \left(8 \Lambda^{2} q^{2}+16 \Lambda  q^{2}+3\right)}}{\Lambda  \left(80 \Lambda^{2} q^{2}+16 \Lambda  q^{2}+12 \Lambda  q +3\right)},
\end{equation}
respectively. Below these temperatures, no black hole can exist. This is what happens in the Hawking-Page transition. For temperatures higher than minimum temperature, two branches are seen. By substituiting the relation (\ref{r01}) or (\ref{r02}) into the Hawking temperature of the system, the minimum temperature value can be estimated. For small horizons, the system is unstable, while for large horizons it is stable. Figure (\ref{fig:temp}) shows the behavior of temperature against the black hole horizon for different values of the involved quantities. As $\Lambda$ increases, the minimum temperature moves upwards.\par
\begin{figure}[h!]
\centering
[a]{\includegraphics[width=7cm]{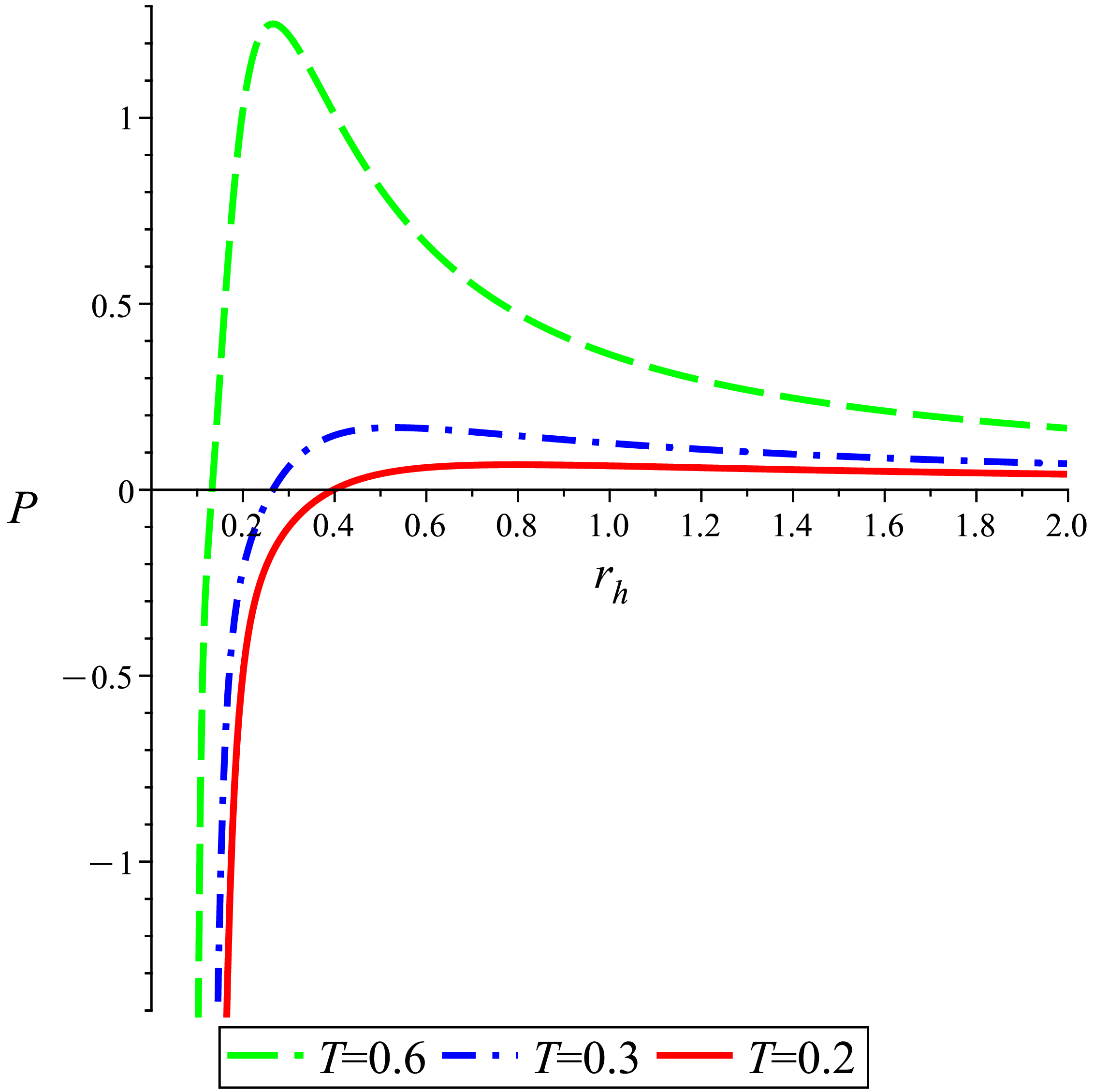}}\quad
[b]{\includegraphics[width=7cm]{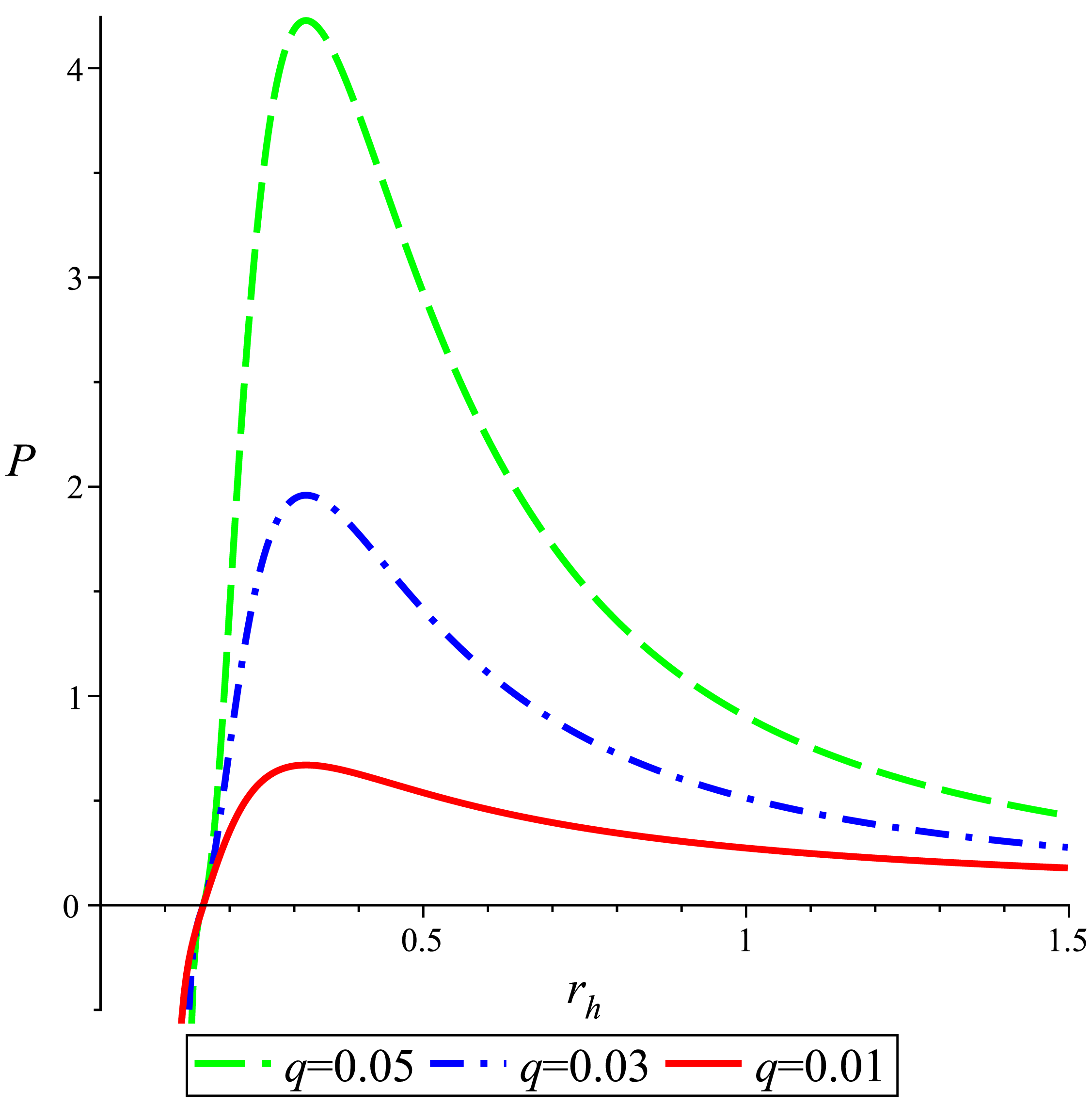}}
\caption{$P-r_h$ diagrams up to the second order of $q$ when \textbf{(a)}: the temperature of the system changes and  \textbf{(b)}: the parameter $q$ changes. . \label{fig:pre}}
\end{figure}

  
The behavior of the pressure against the horizon for different values of temperature and parameter $q$ is seen in Fig. (\ref{fig:pre}). Van der Waal's behavior is not seen in these diagrams. In other words, the conditions
\begin{equation}
  \frac{\partial P}{\partial r_h}|_{T=T_c,r_h=r_c}=0, \qquad  \frac{\partial^2 P}{\partial^2 r_h}|_{T=T_c,r_h=r_c}=0,\nonumber
  \end{equation}
  or
  \begin{equation}
  \frac{\partial T}{\partial r_h}|_{P=P_c,r_h=r_c}=0, \qquad  \frac{\partial^2 T}{\partial^2 r_h}|_{P=P_c,r_h=r_c}=0,\nonumber
  \end{equation}
  are not satisfied in this model. 
Nevertheless, there is a possibility of observing the first and second types of phase transitions in this model. For this purpose, we study the behavior of thermodynamic potentials like entropy and Gibbs free energy to see the signs of the first order phase transition. In order to see the signs of a second order phase transition, the functions like the heat capacity or isothermal compressibility of the system, which are proportional to the second derivative of the Gibbs function, are studied. Before that, let us see the behavior of the entropy versus the system temperature.
\begin{figure}[h!]
\centering
[a]{\includegraphics[width=7cm]{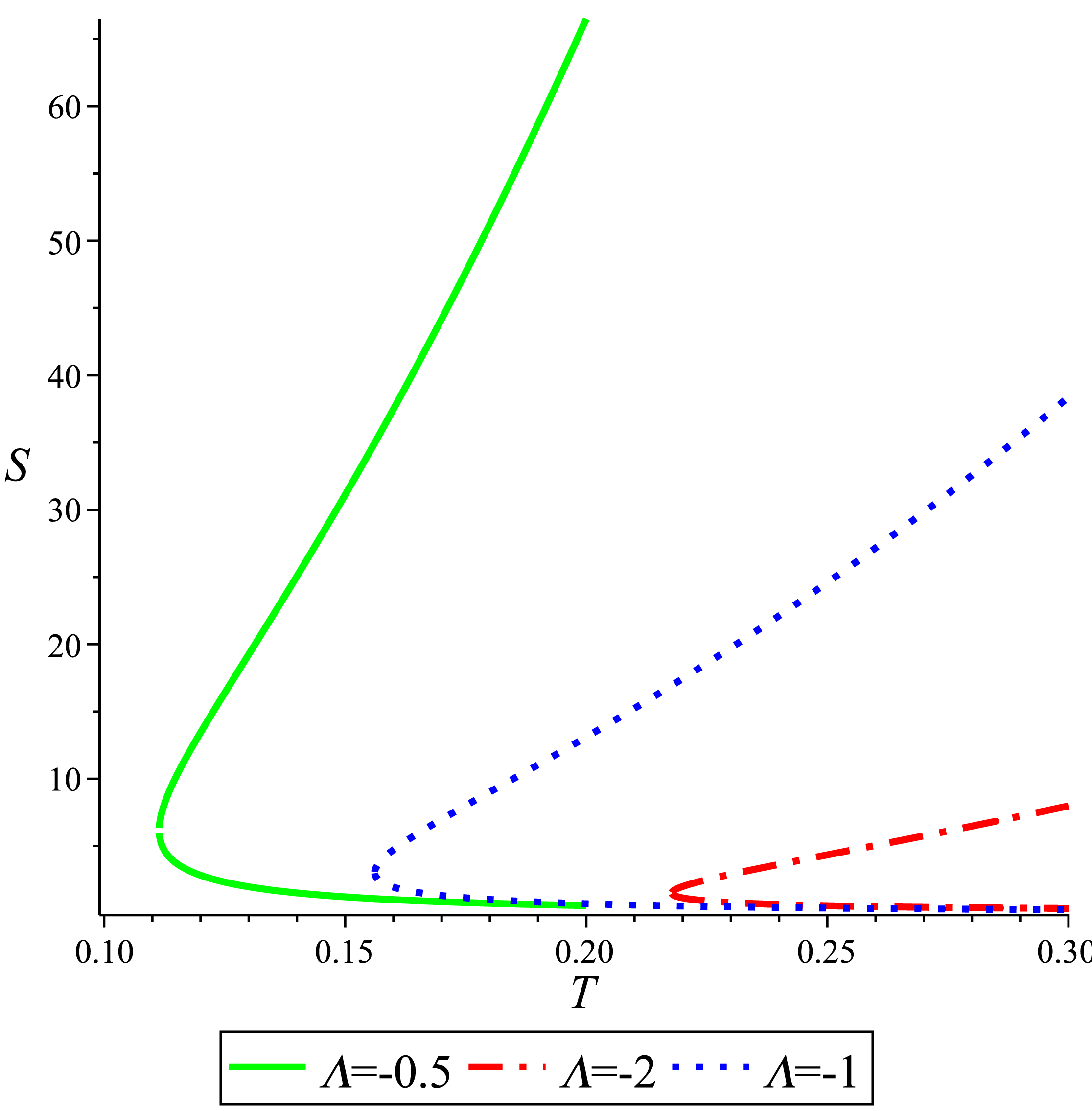}}\quad
[b]{\includegraphics[width=7cm]{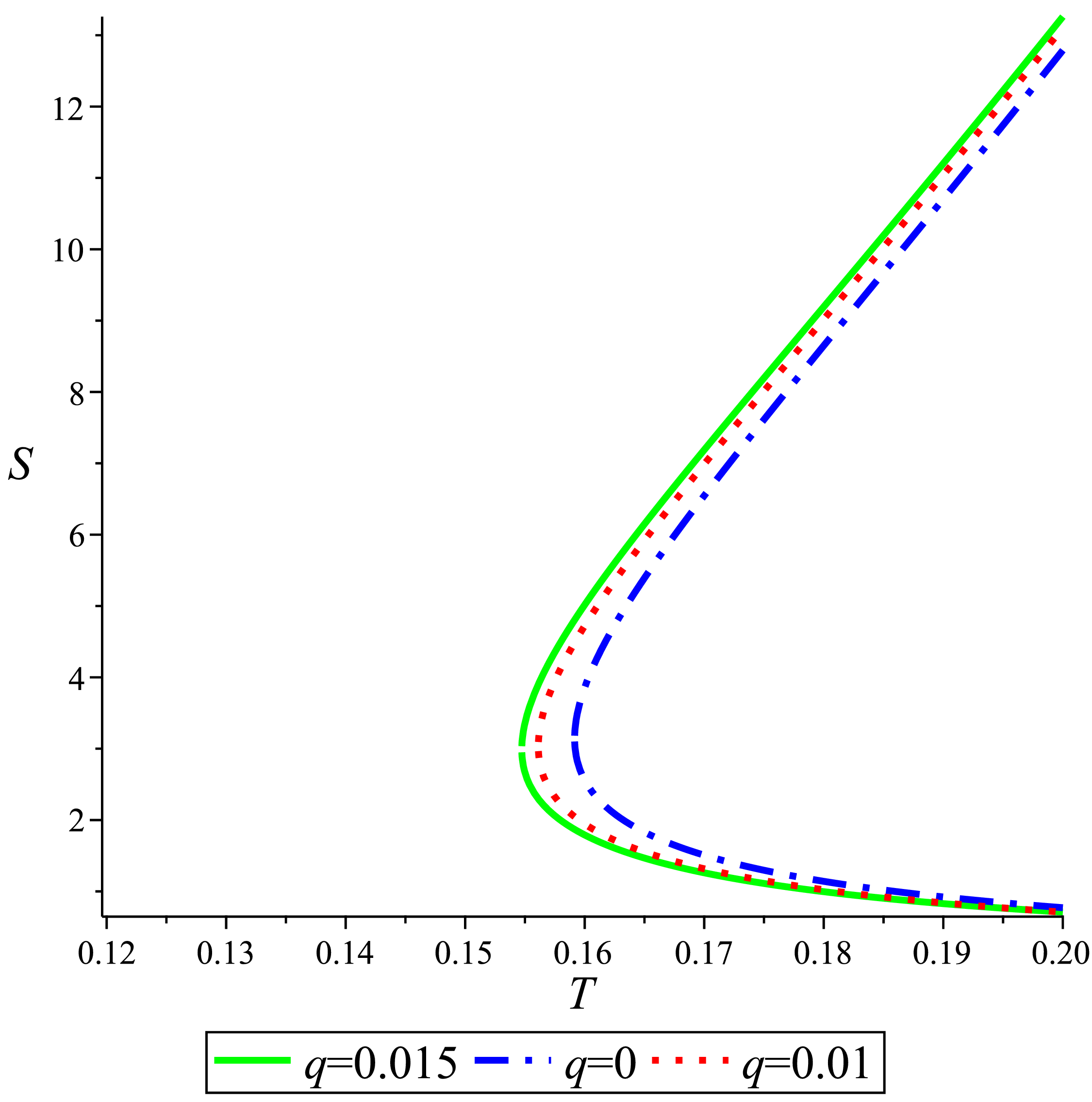}}
\caption{The diagrams of entropy versus temperature when \textbf{(a)}: the value of $\Lambda$ changes, \textbf{(b)}: the value of $q$ changes. A sudden change in the diagrams at a certain point can refer to a phase transition  between unstable and stable black holes. \label{fig:entropy3}}
\end{figure}	
Fig. (\ref{fig:entropy3}) shows the behavior of entropy against the temperature of the system. Phase transitions occur at critical points where the entropy or temperature exhibits sudden changes. The figure shows a transition between large and small black holes. By examining the entropy for several values of parameters $\Lambda$ and $q$, we realized that this is not a stable system in general. The only difference in the diagrams is that the minimum value of the temperature at which the phase transition occurs is shifted to different values of the parameters. This is well illustrated in Fig. (\ref{fig:entropy3}). The slope of the diagrams changes abruptly, indicating a phase transition.\par
\begin{figure}[h!]
\centering
[a]{\includegraphics[width=7cm]{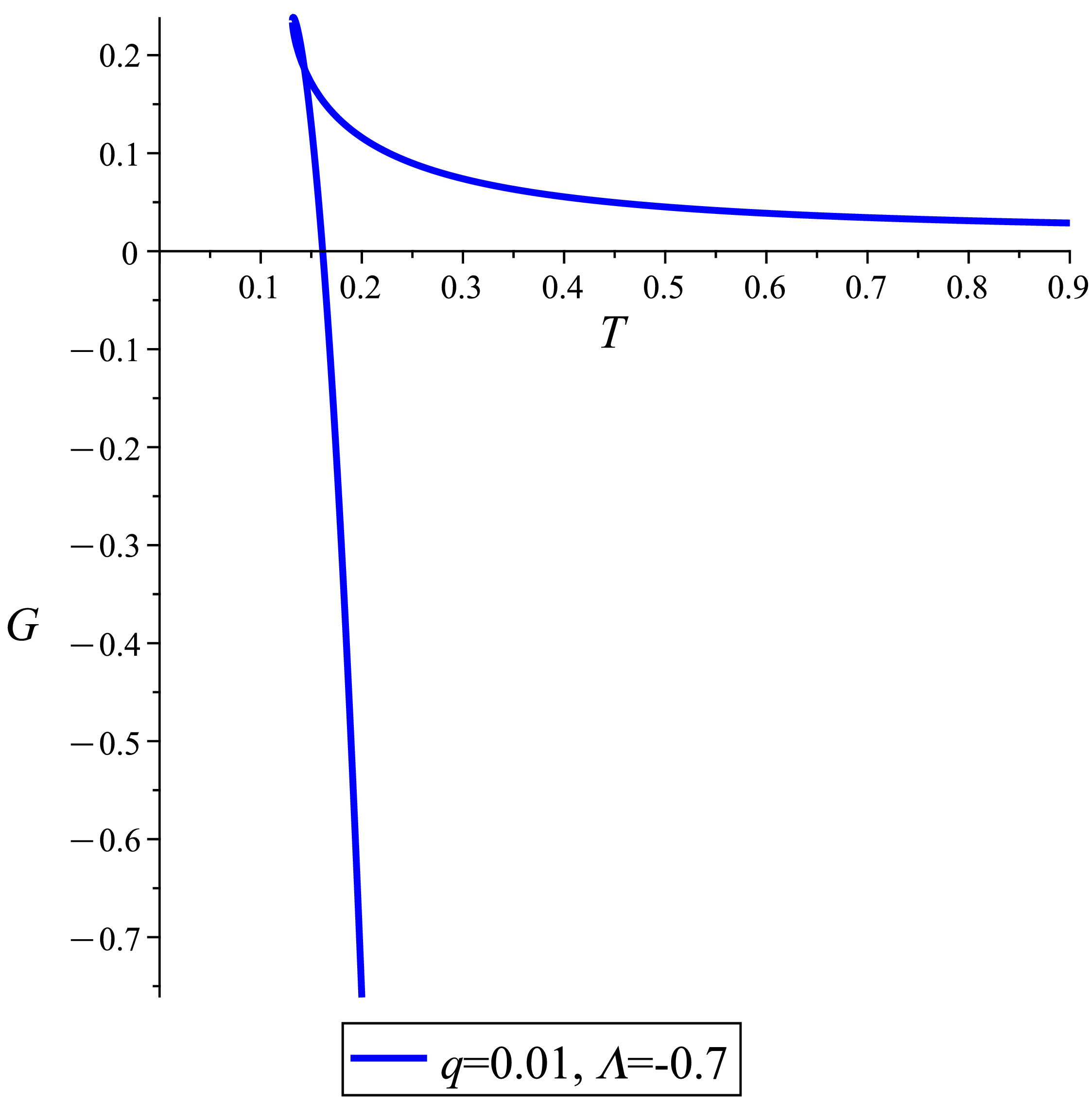}}\quad
[b]{\includegraphics[width=7cm]{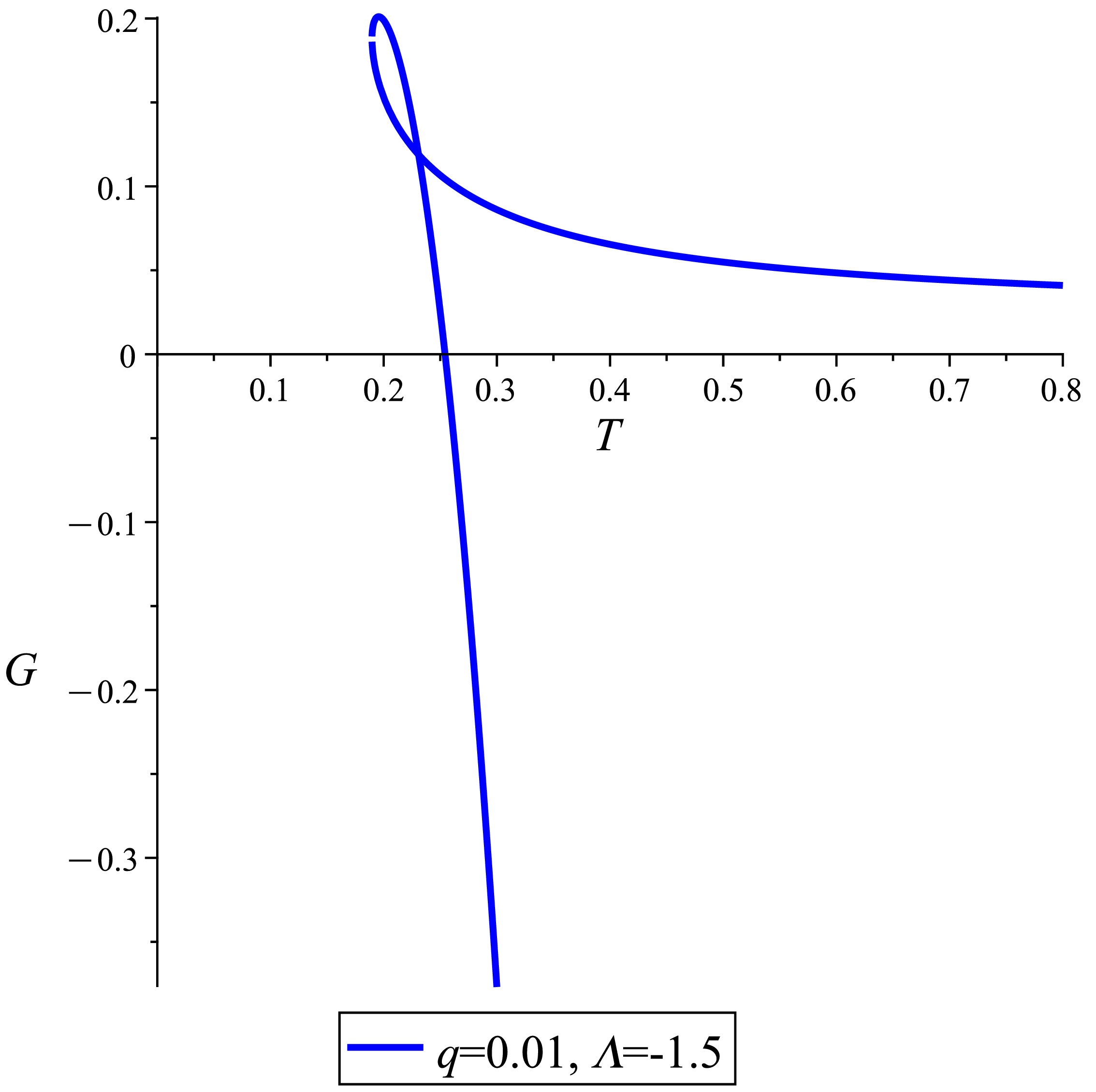}}\quad
[c]{\includegraphics[width=7cm]{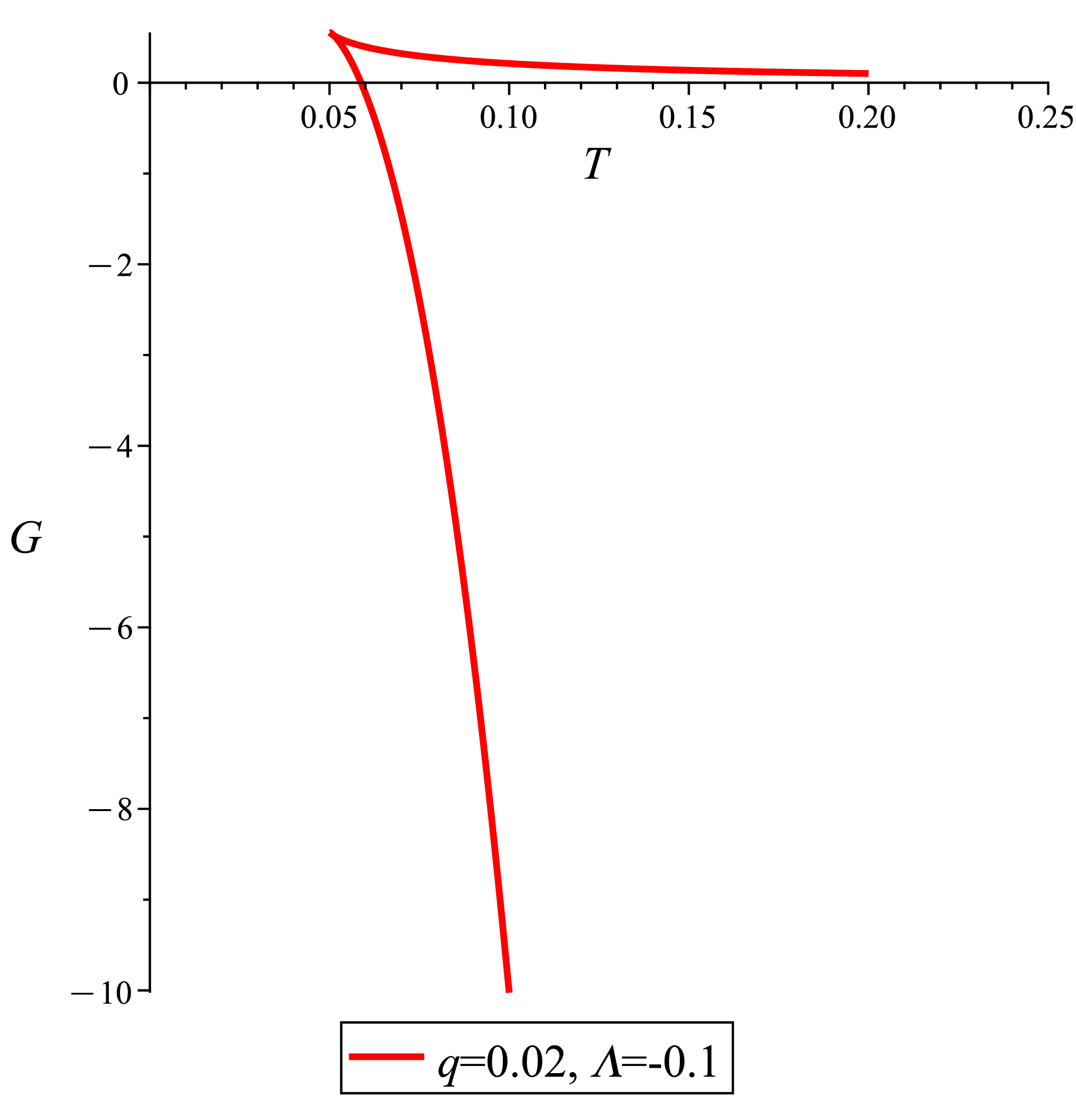}}\quad
[d]{\includegraphics[width=7cm]{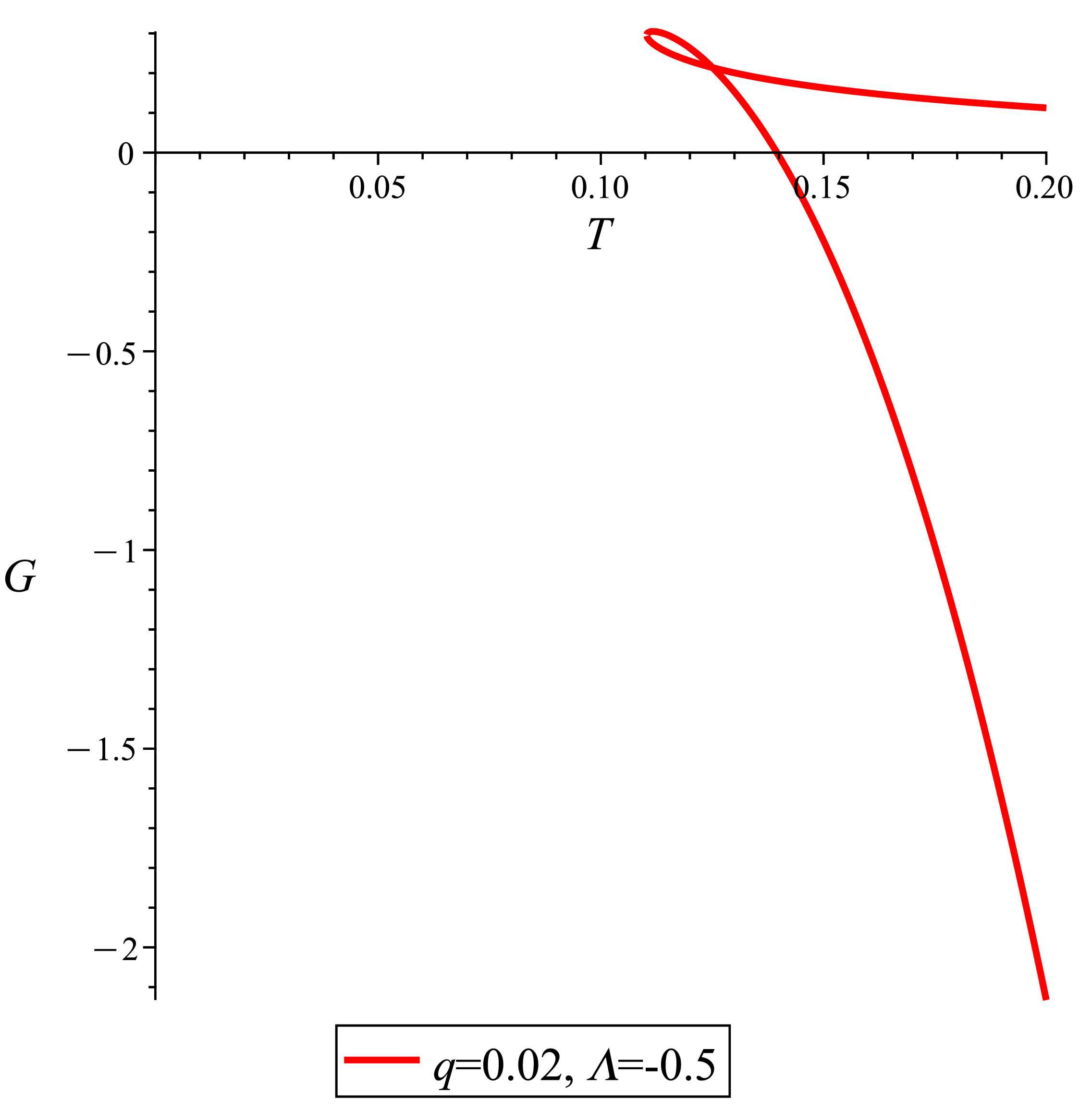}}
\caption{$G-T$ diagrams for the different values of $\Lambda$ and $q$ ; \textbf{(a)}: $q=0.01, \Lambda=-0.7$, \textbf{(b)}: $q=0.01, \Lambda=-1.5$, \textbf{(c)}: $q=0.02, \Lambda=-0.1$ and \textbf{(d)}: $q=0.02, \Lambda=-0.5$. The swallow-tail shapes of these diagrams and their behavior are similar to the Gibbs free energy diagrams of the Hawking-Page transition which indicate a first-order phase transition between small and large black holes. \label{fig:Gibbs2}}
\end{figure}

Now, the Gibbs free energy of this model, which  can give us more detailed information about the type of phase transition, is studied. The Gibbs free energy in terms of the black hole horizon has the following form:
\begin{equation}\label{gr}
  \begin{split}
 G&=M-TS=\frac{r_{h}}{4}+\frac{\Lambda  r_{h}^{3}}{12}+\Big(-2 r_{h} \Lambda +\frac{7}{3} \Lambda^{2} r_{h}^{3}\Big) q\\
 &+\Big(\frac{70 r_{h}^{3} \Lambda^{3}}{3}+\frac{4 \Lambda^{2} r_{h}^{3}}{3}-\frac{40 r_{h} \Lambda^{2}}{3}-\frac{20 r_{h} \Lambda}{3}-\frac{16 r_{h}^{3} \Lambda^{3} \ln \! \left(r_{h}\right)}{3}\\
 &-\frac{8 \Lambda^{2} r_{h}^{3} \ln \! \left(r_{h}\right)}{3}-8 r_{h} \Lambda^{2} \ln \! \left(r_{h}\right)+8 r_{h} \Lambda  \ln \! \left(r_{h}\right)\Big) q^{2}\\
& +\Big(\frac{688 r_{h}^{3} \Lambda^{4}}{3}+\frac{40 r_{h}^{3} \Lambda^{3}}{3}-136 r_{h} \Lambda^{3}-\frac{32 r_{h} \Lambda^{2}}{3}+\frac{32 \Lambda^{2}}{3 r_{h}}-\frac{8 \Lambda}{3 r_{h}}\Big) q^{3}+O(q^4).
  \end{split}
  \end{equation}
By using the relations (\ref{Temp}) and (\ref{gr}) one can obtain a relation for the Gibbs free energy in terms of $\Lambda$ (or the pressure $P$) and the temperature of the black hole. The investigation of the Gibbs function versus temperature for different values of $\Lambda$ and expansion coefficient $q$, leads to the diagrams of Fig. (\ref{fig:Gibbs2}).
In these diagrams, the swallow-tail shapes appear, but they are not exactly like the swallow-tail shapes of a Van der Waals fluid. Nevertheless, these diagrams confirm a first-order phase transition. These diagrams are more similar to the Hawking-Page phase transition diagrams. Where the Gibbs free energy diagrams intersect the horizontal axis($G=0$), it defines the Hawking-Page temperature in this model.\par 
Now, we deal with the heat capacity which is proportional to  the second derivative of the Gibbs free energy and can show the stability or instability of a system. The heat capacity at constant pressure up to the second order of $q$ takes the form
\begin{equation}\label{cr}
  \begin{split}
 &C_P=\frac{\partial H}{\partial r_h} \left(\frac{\partial T}{\partial r_h}\right)^{-1}=\frac{2 \left(r_{h}^{2} \Lambda -1\right) \pi  r_{h}^{2}}{r_{h}^{2} \Lambda +1}+\frac{16 r_{h}^{4} \Lambda^{2} \pi  q}{\left(r_{h}^{2} \Lambda +1\right)^{2}}\\
 &+\frac{64 \Lambda  \left(\left(-\frac{1}{2}+r_{h}^{2} \Lambda^{2}+\frac{\left(r_{h}^{2}+1\right) \Lambda}{2}\right) \left(r_{h}^{2} \Lambda +1\right)^{2} \ln \! \left(r_{h}\right)+\frac{\Lambda^{4} r_{h}^{6}}{2}+2 r_{h}^{4} \Lambda^{3}+3 r_{h}^{2} \Lambda^{2}+\frac{\Lambda}{2}\right) r_{h}^{2} \pi  q^{2}}{\left(r_{h}^{2} \Lambda +1\right)^{3}}\\
 &+O(q^3).
  \end{split}
  \end{equation}
The denominator of (\ref{cr}) shows that the heat capacity diverges only at one point. The divergence point changes with changes in the pressure of the system. The interesting point is that the position of the divergence point does not change with the changes of the  parameter $q$. According to  relation (\ref{cr}), the divergence point occurs at horizon $r_d=\frac{1}{\sqrt{-\Lambda}}$. The heat capacity diagrams show that we have only two unstable and stable phases in this model, which correspond to small and large black holes. There is no intermediate phase in this model.
\begin{figure}[h!]
\centering
[a]{\includegraphics[width=7cm]{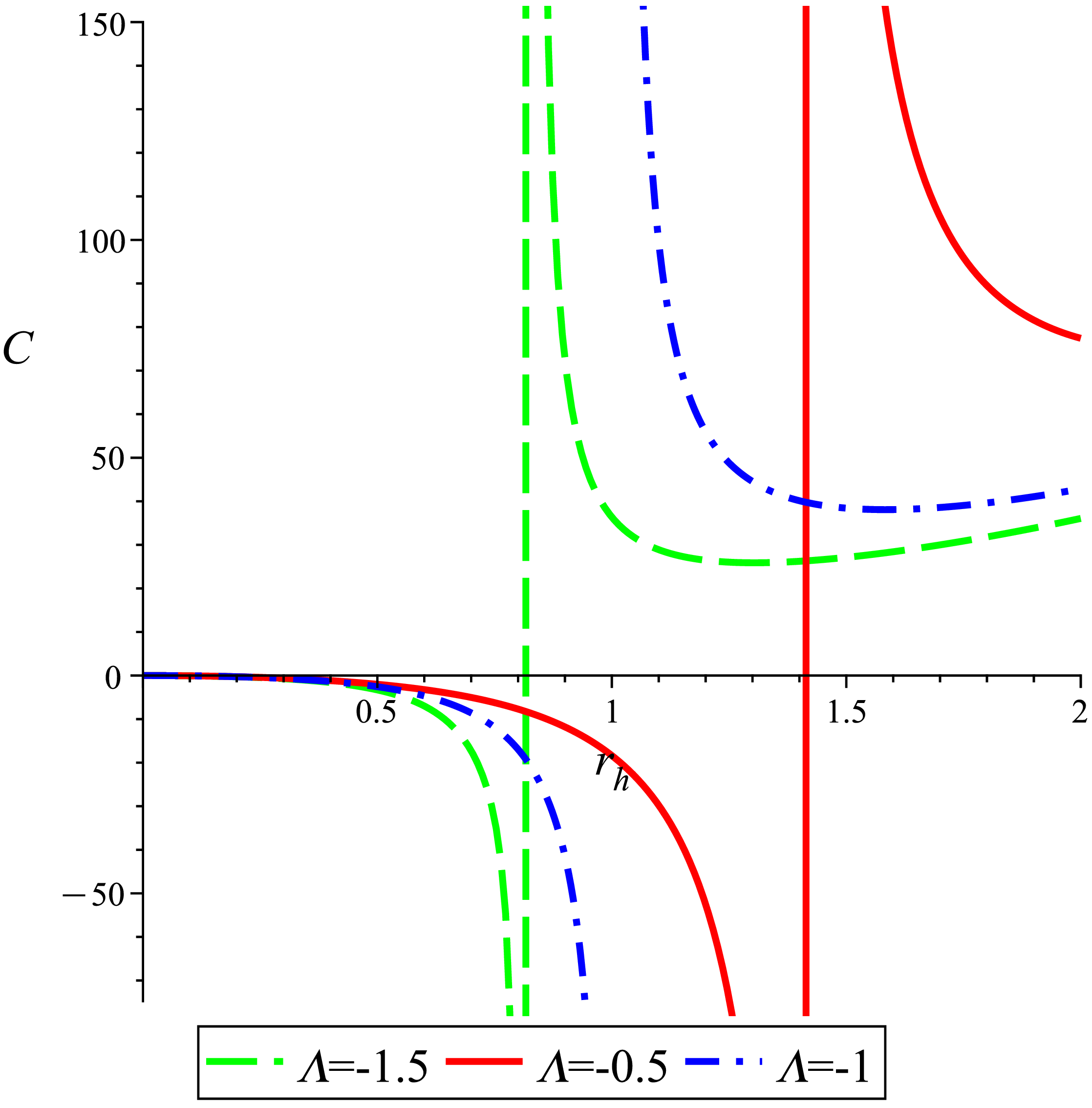}}\quad
[b]{\includegraphics[width=7cm]{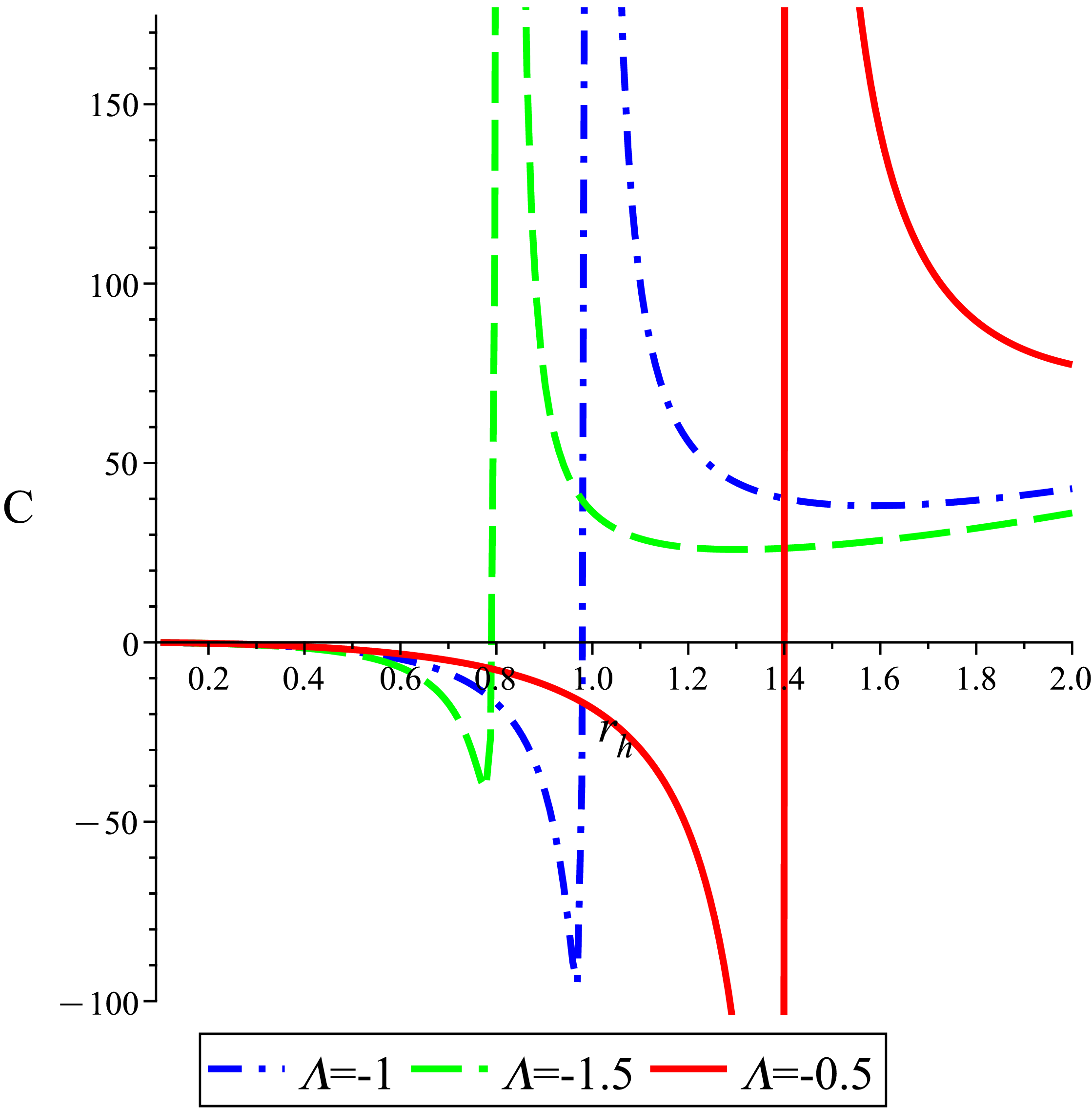}}
\caption{The diagrams of the heat capacity versus the black hole horizon for different values of $\Lambda$ and $q=0.01$; \textbf{(a)}: up to the second order of $q$ and  \textbf{(b)}: up to the third order of $q$. By increasing the value of $\Lambda$, the heat capacity diverges at smaller horizons. \label{fig:capacity1}}
\end{figure}
  \begin{figure}[h!]
\centering
[a]{\includegraphics[width=7cm]{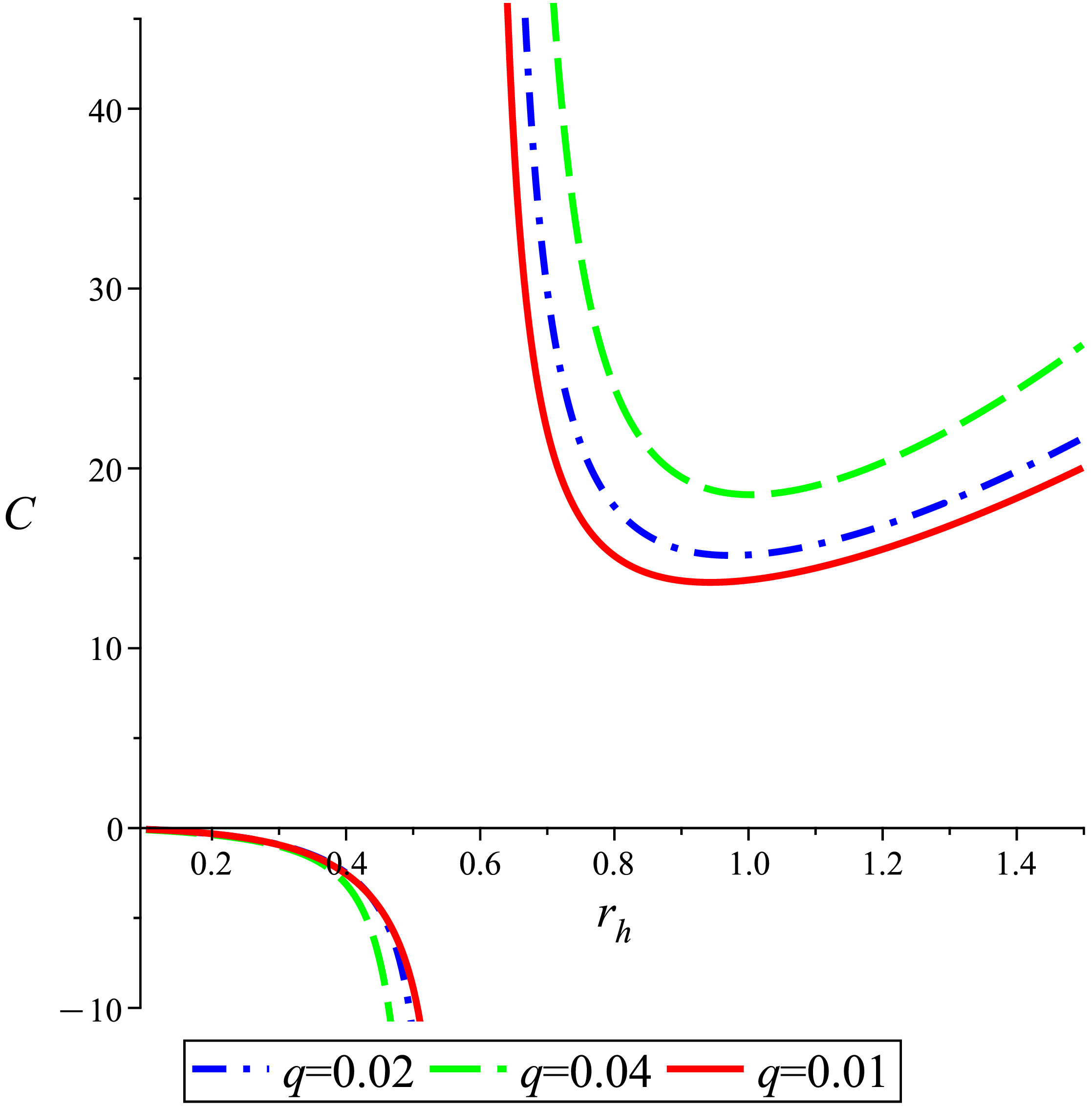}}\quad
[b]{\includegraphics[width=7cm]{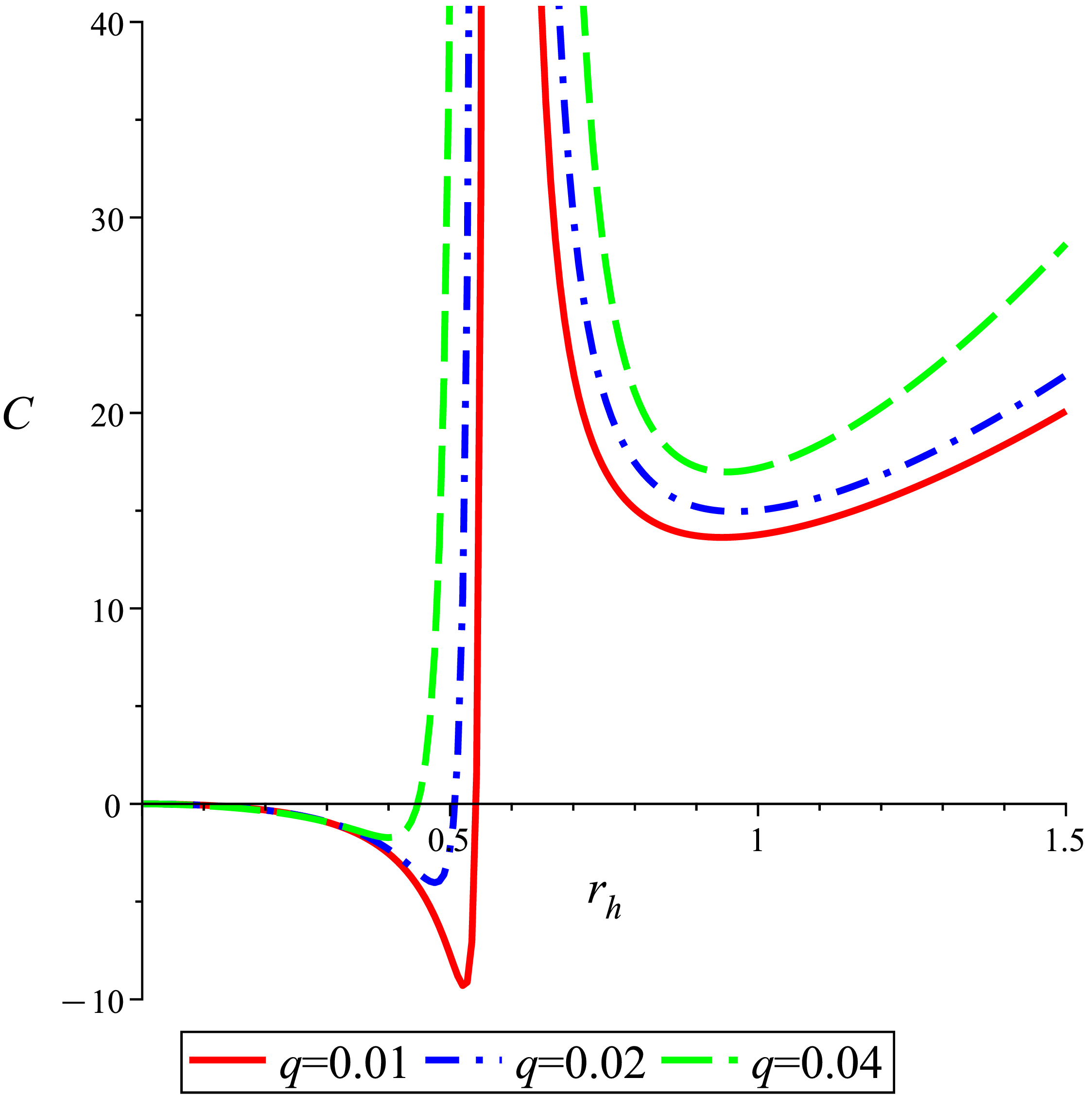}}
\caption{The diagrams of the heat capacity versus the black hole horizon for different values of $q$ and $\Lambda=-3$; \textbf{(a)}: up to the second order of $q$ and  \textbf{(b)}: up to the third order of $q$. The divergence point is independent of the value of the parameter $q$. \label{fig:capacity2}}
\end{figure}
\begin{figure}[h!]
\centering
[a]{\includegraphics[width=7cm]{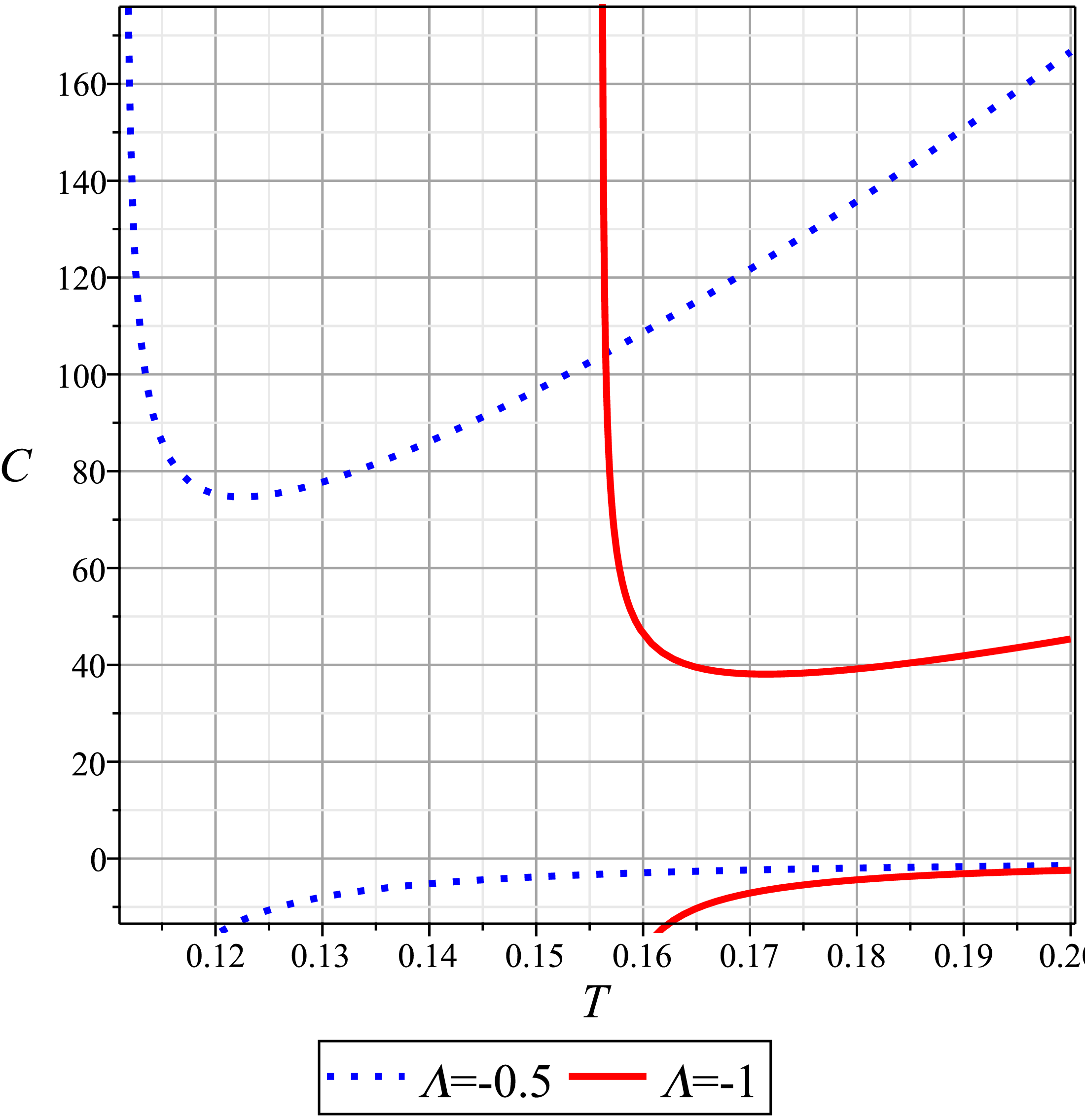}}\quad
[b]{\includegraphics[width=7cm]{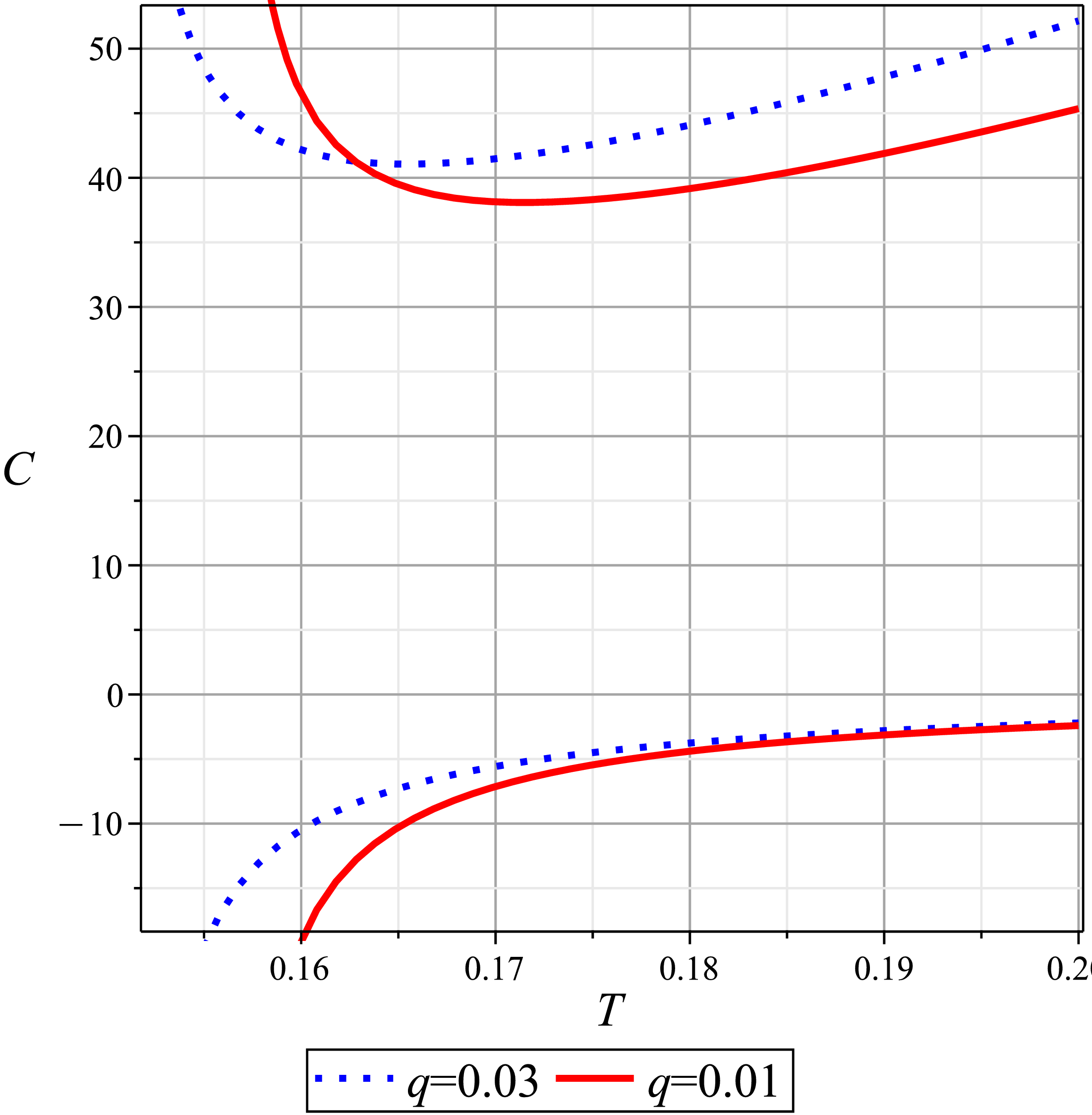}}
\caption{The diagrams of the heat capacity versus the black hole temperature up to the second order of $q$ when, \textbf{(a)}: $\Lambda$ changes and  \textbf{(b)}: the parameter $q$ changes. Different branches above and bellow the temperature axis show the stable and unstable  black hole states.\label{fig:capacity3}}
\end{figure}
Figures (\ref{fig:capacity1}) and (\ref{fig:capacity2}) represent the behavior of the heat capacity against the horizon of the black hole. As it is obvious, a transition between an unstable small black hole ($C_P<0$) to a large stable black hole ($C_P>0$) occurs.  These divergence points in the diagrams of heat capacity can be signs of a second order phase transition. Because in the second order phase transition, the second derivative of the Gibbs free energy is discontinues. And the heat capacity is proportional to the second derivative of the Gibbs free energy. In the first order phase transition, the entropy which is proportional to the first derivative of the Gibbs free energy  has a jump or varies suddenly at a point. Since. By increasing the value of $\Lambda$, the phase transition takes place at smaller horizons. The right panels of figures (\ref{fig:capacity1}) and (\ref{fig:capacity2}) correspond to the heat capacity up to the second order of $q$. In these cases, the heat capacity can have a root near the divergence points. In general, the roots indicate the phase transitions between stable and unstable phases in black holes. The divergence points are associated with critical behavior and phase transitions. \par 
Figure (\ref{fig:capacity3}) shows the behavior of heat capacity versus temperature. Two distinct branches in each plot refer to the two unstable(negative capacity) and stable(positive capacity) phases of the black hole. By increasing the value of $\Lambda$, the diagram of the heat capacity diverges at higher temperatures. While, by increasing the value of $q$, the heat capacity diverges at lower temperatures.
%--------------------------------------------------------------------------
 \section{Conclusion}
A general exponential modification for an AdS black hole action was considered. The solution for the metric function $f(r)$ was obtained through a perturbative method up to the third order of the perturbation parameter. The behavior of the metric function for different values of parameters was depicted. Then, the thermodynamic behavior and phase transition of the black hole was studied. Through the thermodynamic study of this model we realized that the behavior of a Van der Waals fluid is not seen in this model. The necessary and important relations for the thermodynamic quantities were derived. The diagrams of the temperature versus the black hole horizon show  behaviors like that of the Hawking-Page temperature diagrams and phase transitions between thermal AdS space and AdS black holes. The Smarr formula for this model was derived up to the linear order in $q$. For this purpose, we had to consider the parameter $q$ as a thermodynamic variable conjugated to the potential $\mathcal{Q}$. The entropy function against the black hole horizon does not behave abnormally up to the third order of $q$, but from the fourth order of $q$, for small horizons, the entropy increases. This cannot be explained classically. In this study, the diagrams of entropy versus temperature, show the sudden change in direction at specific points. This indicates a phase transition between the unstable and stable black hole states. The investigation of the Gibbs free energy versus temperature at constant pressure  showed that it is possible to have a first-order phase transition in this model.In fact, the Gibbs free energy diagrams became very similar to the Gibbs diagrams of Hawking-Page transition. And as we know, the Hawking-Page transition is a transition of the first type. Also, the divergence points of the heat capacity signal a second order phase transition. The divergence points of the heat capacity depend only on $\Lambda$ and do not depend on the value of the parameter $q$. Perhaps, by adding the matter field to the gravitational action, effects similar to the van der Waals fluid can be observed, which can be investigated in a separate work.\par 
According to these results and the dictionary of AdS/CFT correspondence, we conclude that this model can be considered for the confinement-deconfinement phase transition and also a phase transition from a normal phase to a superconducting phase in the dual conformal field on the boundary of the AdS space.

%--------------------------------------------------------------------------
%\vspace{1cm}
%\noindent {\large {\bf Acknowledgment} } We would like to thank Shahrokh Parvizi, Matteo Baggioli, Mojtaba Shahbazi and Komeil Babaei for useful comments and suggestions. We also thank the referees of CJP for the valuable comments which helped us to improve the manuscript.
%--------------------------------------------------------------------------

\vspace{1cm}
\noindent {\large {\bf Data Availability } } Data generated or analyzed during this study are provided in full within the published article.
%--------------------------------------------------------------------------

\vspace{1cm}
\noindent {\large {\bf Competing interests }  The authors declare that there are no competing interests.
%--------------------------------------------------------------------------

\end{document}